\documentclass[10pt]{wlscirep}
\usepackage[utf8]{inputenc}
\usepackage[T1]{fontenc}
\renewcommand*{\thesection}{\arabic{section}}
\usepackage{amsmath,graphicx,float}

\usepackage{bm}
\usepackage{amssymb}
\usepackage{graphicx,xcolor}
\usepackage{csquotes}
\usepackage{bm}
\usepackage{amssymb}
\usepackage{graphicx,xcolor}
\usepackage{yfonts}
\usepackage{amsmath,graphicx}
\usepackage{epigraph}
\usepackage{csquotes}
\usepackage{moresize}
\begin{document}
\def\thesection{\Roman{section}}
\def\tr{\rm{Tr}}
\def\la{{\langle}}
\def\ra{{\rangle}}
\def\a{{\alpha}}
\def\e{\enquote}
\def\q{\quad}
\def\w{\tilde{W}}
\def\t{\tilde{t}}
\def\ett{\tilde{\eta}}
\def\x{\tilde{x}}
\def\k{\tilde{k}}
\def\aa{\underline{\alpha}}
\def\mm{\underline{\mu}}
\def\xx{\overline{x}}
\def\m{\mu}
\def\pp{\underline{p}}
\def\tt{\underline{t}}
\def\xx{\underline{x}}
\def\uu{\underline{U_0}}
\def\a{\hat{A}}
\def\h{\hat{H}}
\def\Res{\text{Res}}
\def\E{\epsilon}
\def\E{\mathcal{E}}
\def\p{\hat{P}}
\def\R{\text{Re}}
\def\Ip{\text{Im}}
\def\u{\hat{U}}
\def\n{\\ \nonumber}
\def\j{\hat{j}}
\def\et{\tilde{\eta}}
\def\g{\hat{G}}
\def\crr{\color{black}}
\def\cbb{\color{black}}
\def\al{\alpha}
\def\vc{\underline{c}}
\def\vf{\underline{f}}

\title{Speed-up and slow-down of a quantum particle}

\author[1]{X. Guti\'errez de la Cal*}
\author[2]{M. Pons}
\author[1,3]{D. Sokolovski}

\affil[1]{Departamento de Qu\'imica-F\'isica, Universidad del Pa\' is Vasco, UPV/EHU, Leioa, Spain}
\affil[2]{Departamento de F\' isica Aplicada, Universidad del Pa\' is Vasco, UPV-EHU, Bilbao, Spain}
\affil[3]{IKERBASQUE, Basque Foundation for Science, E-48011 Bilbao, Spain}


\begin{abstract}
\noindent
We study {\crr non-relativistic} propagation of Gaussian wave packets in one-dimensional Eckart potential, a barrier, or a well.
In  the picture used, the transmitted wave packet results from interference
between the copies of the freely propagating state with different spatial shifts (delays), $x'$, induced by the scattering potential.
The Uncertainty Principle precludes relating the particle's final position to the delay experienced in the potential, except
in the classical limit.
Beyond this limit, even defining an effective range of the delay is shown to be an impracticable task, owing to the
oscillatory nature of the corresponding amplitude distribution.
Our examples include the classically allowed case, semiclassical tunnelling, delays induced in the presence of a virtual
state, and scattering by a low barrier. The properties of the amplitude distribution of the delays, and its pole representation
are studied in detail.
\end{abstract}

\flushbottom
\maketitle
%
%
\thispagestyle{empty}

\section{ Introduction}
A classical particle, crossing in one dimension a short-range potential $V(x)$,
goes faster over a well, $V(x)<0$, and slower over a barrier, $V(x)>0$.
Once it  has left the potential, this can be checked either by evaluating  the distance $x'$ separating the particle
from its freely moving counterpart at a given time $t$, or by measuring the time
interval $\tau$ between the two arrivals at a fixed detector.
The reason is simple. The particle's velocity inside the potential either exceeds that of the free motion,
or is reduced, and $\tau=x'/v_0$, where $v_0$ is the speed of free motion.
\newline
This explanation relies on the existence of a classical trajectory, and can no longer be used in the quantum case.
There, a particle described by a wave packet may tunnel across the barrier even if its energy is smaller than
the barrier's height.  Attempts to ascribe to tunnelling a single duration $\tau$ began with McColl's observation that
\e{there is no appreciable delay in transmission of the  packet through the barrier} \cite{McColl}, and continue to date,
encouraged by the recent progress in atto-second experimental techniques \cite{{Swiss},{att1},{att2}}.
The discussion has recently become a dispute between those who think that $\tau$ should be zero \cite{att1},
and those believing that it should have a non-zero value \cite{{att2},{Stein}}.
\newline
Before estimating the value of a commodity, it may be useful to enquire whether the commodity in question does indeed exist.
This is particularly true in the case of tunnelling, where the existence of a well defined \e{tunnelling time} can be shown to contradict
the Uncertainty Principle \cite{{DSE1},{DSE2}}.
Recently, there has been renewed interest in the tunnelling time problem (see \cite{{Pollak1},{Pollak2},{Diener},{Petersen1},{Petersen2},{Dumont2},{Rivlin}}).
Although tunnelling is, without a doubt, the most interesting example, there are many other situations in which the classical
analysis should fail. For example, such would be the case of a slow particle moving across a shallow well,
supporting only few bound states, or passing over a low potential barrier. In brief, it would be helpful to have a general approach
to quantum scattering, setting clear limits on what can and what cannot be asked of a quantum particle.
Such an approach was outlined, and applied to wave packet tunnelling, in \cite{DSE0}. In this paper,
we discuss  a more general application of the method to the transmission across an Eckart potential \cite{Land}.
Unlike the rectangular barriers and wells, often discussed in a similar context, Eckart's potential  is amenable to standard
semiclassical treatment \cite{Land}. The corresponding transmission amplitude, $T(p,V)$,  is known analytically,
as well as the positions and residues of  its poles in the complex momentum plane. All this makes this potential an ideal candidate
for our demonstration.
\newline
The rest of the paper is organised as follows.
In Section II we briefly discuss
 Eckart's potential and Gaussian wave packets, used throughout the rest of the paper.
Section III discusses the classical limit for passing over an Eckart's well, or barrier.
In Section IV we analyse the semiclassical limit of a tunnelling transmission.
In Section V a change to the coordinate representation allows us to describe a potential
as a kind of an \e{interferometer} which splits the initial wave packet into the
components with different spatial delays, recombined later to produce the transmitted state.
In Section VI we introduce a pole representation for the amplitude distribution of the delays.
The cases of an Eckart's barrier and an Eckart's well are analysed separately in Sections VII and VIII,
respectively.
In Section IX we ask whether one can define an \e{effective range} of the delays, and answer the question in the
negative. Section X discusses the relation between special delays and measurable time intervals.
In Section XI we relate the \e{phase time} to the displacement of the centre of mass of a wave packet, broad in the
coordinate space. The delay experienced by a slow particle in a shallow Eckart well  is discussed in Section XII.
The case of a low Eckart barrier is analysed in Section XIII.
Our conclusions are in Section XIV.


\section{ Gaussian wave packets in an Eckart's potential}
\begin{figure}[h]
\centering
\includegraphics[width=0.5\linewidth]{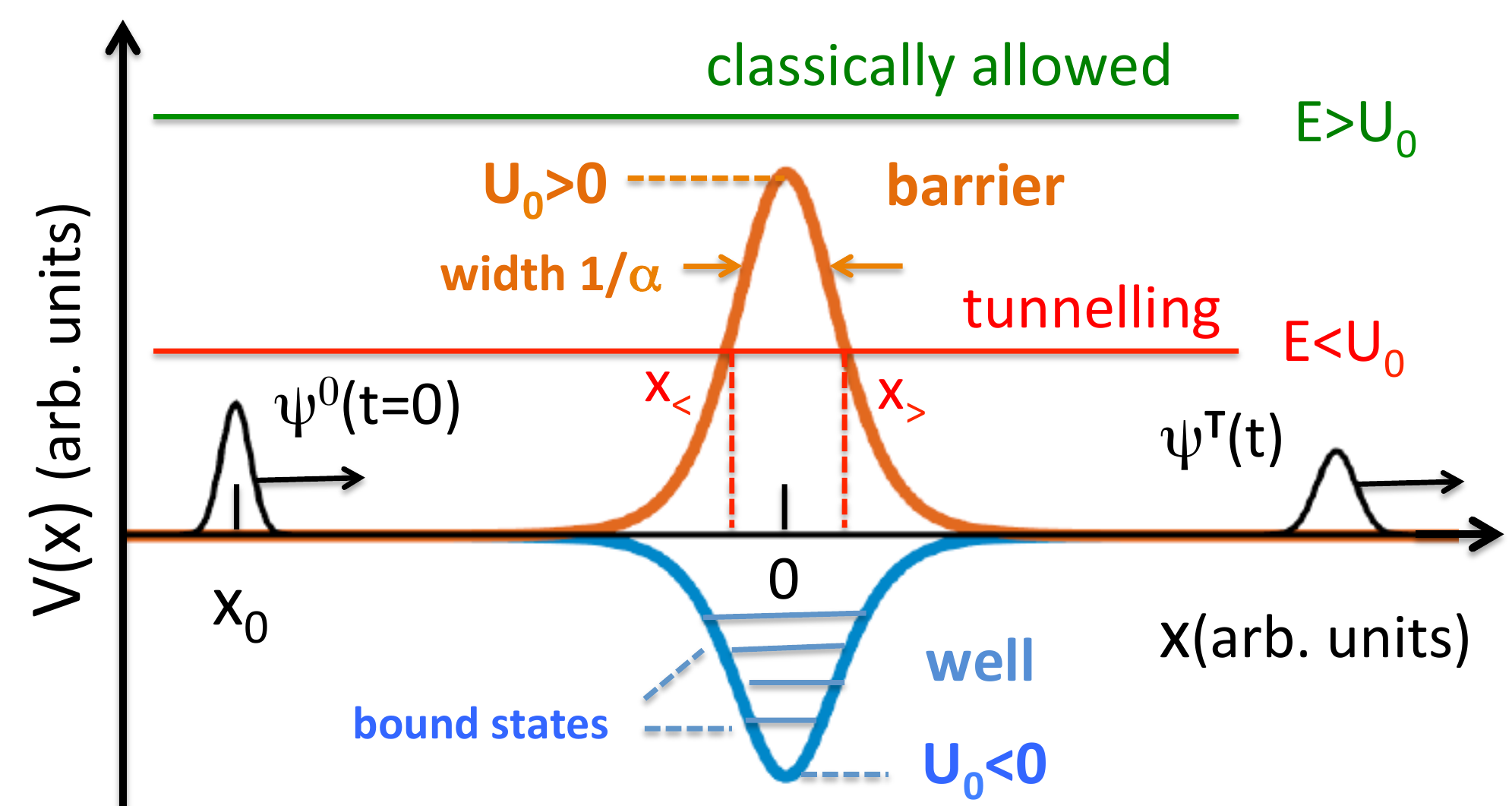}
\caption {Eckart's potentials, a barrier and a well. Also shown schematically
are the initial, $\psi^0(x,t=0)$, and the transmitted, $\psi^T(x,t)$, wave packets. }
\label{0}
\end{figure}

We consider, in one dimension, scattering
of a non-relativistic wave packet incident  from the left on a smooth potential  $V(x)$, $V(x)\to 0$ for $x\to \pm \infty$, see \textbf{Figure \ref{0}}.

For the initial wave packet we have [$E(p)=p^2/2\mu$, where $\mu$ is the particle's mass, and $\hbar=1$ is used]
\begin{align}\label{-1}
\psi^0(x,t) =\int A(p-p_0) \exp[ipx -iE(p)t] dp.
\end{align}
 Its transmitted part is given by
\begin{equation}\label{-2}
\psi^T(x,t) =\int T(p,V) A(p-p_0) \exp[ipx -iE(p)t] dp,\q
\end{equation}
where $T(p,V)$ is the transmission amplitude.
At a time $t$, large enough for the scattering to be completed, we will compare the positions of the transmitted
wave packet with that of a freely propagating one, in order
to determine whether the potential delays the particle or makes it, in some sense, go faster.
In particular, we will consider an Eckart potential \cite{Land}
\begin{equation}\label{-3}
V(x)=\frac{U_0}{\cosh^2(\alpha x)},
\end{equation}
a well, or a barrier, depending on the sign of $U_0$.
For such a potential the transmission amplitude is well known to be \cite{Land}
\begin{equation}\label{-4}
T(p,V)= \frac{\Gamma(-ip/\alpha-s)\Gamma(-ip/\alpha+s+1)}{\Gamma(-ip/\alpha)\Gamma(1-ip/\alpha)},
\end{equation}
where $\Gamma(z)$ is the Gamma function \cite{Abr}, and
\begin{equation}\label{-5}
s\equiv 2^{-1 }\left [-1+\sqrt{1-\frac{8\mu U_0}{\alpha^2}}\right ].
\end{equation}
The transmission amplitude has the usual property \cite{Baz}
(a star denotes complex conjugation)
\begin{equation}\label{-5a}
T(-p^*,V)= T^*(p,V),
\end{equation}
which can also be obtained directly from (\ref{-4}).
\newline
We will be interested in Gaussian wave packets,  and choose the momentum distribution in equation  (\ref{-1})
to be
\begin{align}\label{-6}
A(p-p_0)&=2^{-1/4}\pi^{-3/4}\Delta p^{-1/2}\times \exp[-(p-p_0)^2/\Delta p^2-i(p-p_0)x_0],
\end{align}
where $x_0 <0$.
With this, at $t=0$, the incident wave packet is a Gaussian state with a mean momentum $p_0$, of a width
$\Delta x=2/\Delta p$, placed at $t=0$ a distance $|x_0|\gg 1/ \alpha$  to the left of the barrier.
In the coordinate representation we, therefore, have
\begin{equation}\label{-4a}
\psi^0(x,t) = \exp[ip_0x-iE(p_0)t]G_0(x,t)
\end{equation}
where the envelope $G_0(x,t)$ is given by equation 
{\crr(S$2$) of the Supplementary Appendix A}.
It will be convenient to measure the distances in the units of the barrier's width $1/\alpha$,
and use dimensionless variables,
\begin{align}\label{-5b}
\mm=1,\q \aa=1,\q \xx=\alpha x,\q \pp=p/\alpha,\nonumber\\  \tt=\alpha^2 t/\m, \q \uu= \m U_0\alpha^2.\q\q\q\q\q
\end{align}
Next we consider the classical limit.
\section{The classical limit. Spatial advances and delays.}
\begin{figure}[h]
\centering
\includegraphics[width=0.5\linewidth]{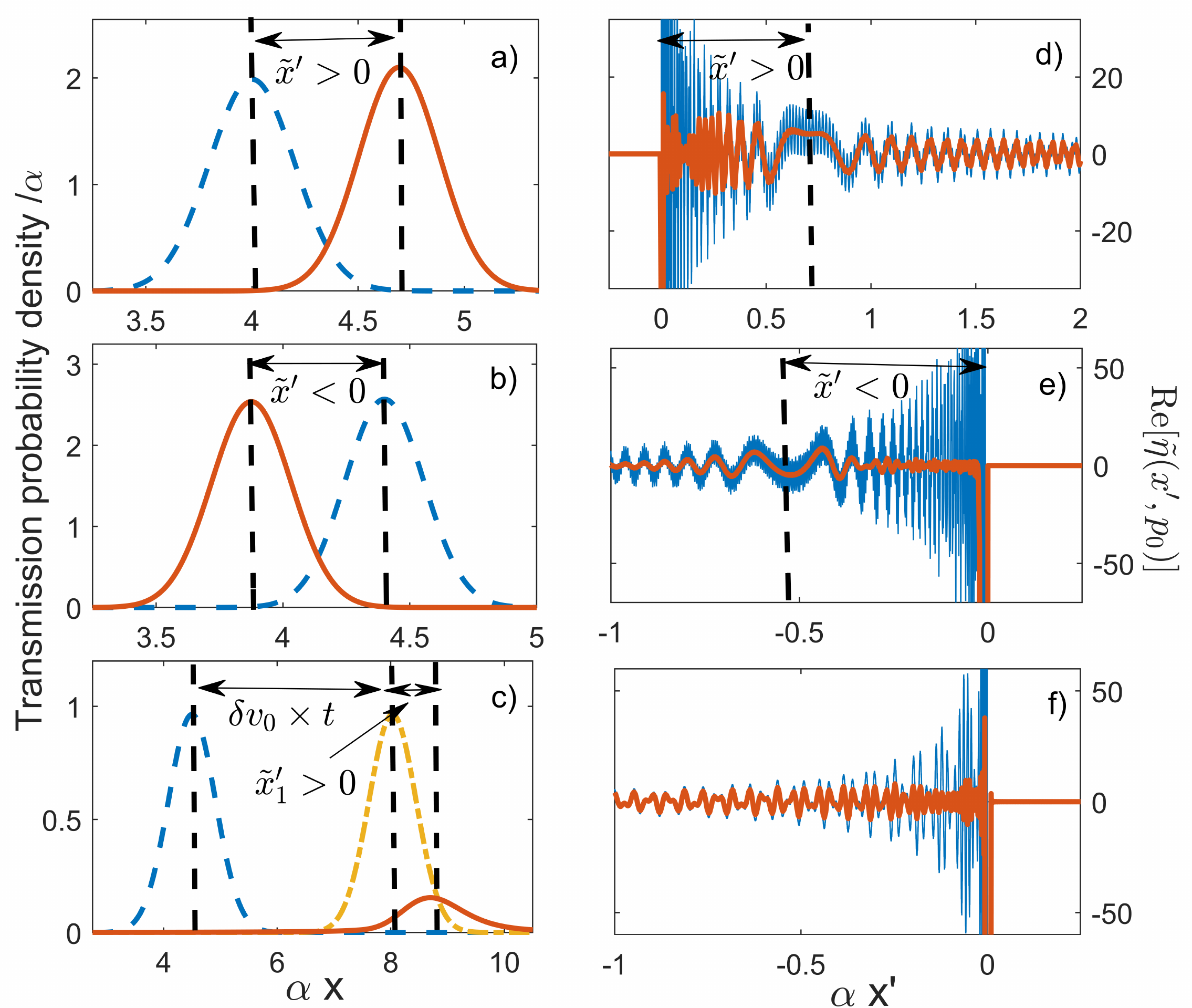}
\caption {{\bf a})  A wave packet is transmitted across an Eckart well,  $\uu=-2\times 10^4$, $\pp_0=200$, $\Delta \pp=6.67 $, $\xx_0=-4$, and
$\tt=0.04$. Also shown by the dashed line is its freely propagating counterpart in equation (\ref{1}).
The spatial advancement calculated
in equation (\ref{a4})  is $\alpha \tilde x'=0.6619$.
\newline
{\bf b}) Same {\bf a}) but for a passage  above an Eckart  barrier,  $\uu=10^5$, $\pp_0=700$,  $\Delta \pp= 6.67 $, $\xx_0=-4$, and
$\tt=0.012$.
The spatial delay calculated
in equation (\ref{a4})  is $\alpha \tilde x'=-0.5392$.
\newline
{\bf c}) Wave packet (multiplied by $z=10^{221}$ for better viewing)  tunnels across an Eckart  barrier,  $\uu=10^4$, $\pp_0=50$,  $\Delta \pp= 3.64 $, $ \xx_0=-4$, and
$\tt=0.17$.
Freely propagating wave packets with mean momenta $p_0$ and $p_0+\delta p_0$,
are shown by dashed and dot-dashed lines, respectively.
The spatial  advancement
in equation (\ref{b2})  is $\alpha \tilde x'=0.8515$.
\newline
{\bf d}), {\bf e}) and {\bf f}) show $\R[\tilde \eta(p_0,x')]$ (solid) and its average,
$\R[y^{-1}\int_{x'}^{x'+y}\tilde \eta(p_0,x'')dx'']$  (thick solid) for the cases in {\bf a}), {\bf b}) and {\bf c}), respectively.
Where it exists, the classical value $\tilde {x}'$ is marked by a vertical dashed line.}
\label{1}
\end{figure}

For a classically allowed motion we have $E(p)>V(x)$, and  the local momentum,
$q(x,p)=\sqrt{p^2-2\mu V(x)}$, is a real positive quantity. The classical requirement,
that the potential should vary slowly on the scale of a local de Broglie wavelength \cite{Land},
now translates into a condition $1/\text {min}[q(x,p)]\ll 1/\alpha$.
Neglecting the small over-barrier reflection, we write the transmission amplitude (\ref{-4}) as \cite{Brink}
\begin{align}\label{a1}
T(p,V)\approx &\exp\left \{i\int_{-\infty}^\infty [q(x,p)-p]dx\right \}
\equiv  \exp[i\Phi(p,V)],  \q\q
\end{align}
and expand the phase $\Phi$ in a Taylor series around the particle's mean momentum $p_0$,
\begin{align}\label{a2}
\Phi(p,V) = &\Phi(p_0,V)-\int_{-\infty}^\infty dx
\left[1-\frac{p_0}{q(x,p_0)}\right ](p-p_0)
+\sum_{n=2}^\infty \partial_p^n \Phi(p_0,V)(p-p_0)^n/n!.
\end{align}
In the classical limit one expects a wave packet of a {size}  $\Delta x \sim 1/\Delta p$, small compared to the
size of the potential $\delta x \sim 1/\alpha$, to be transmitted without distortion, and experience a delay or
an advancement, depending on whether $V(x)$ is a barrier or a well.
\newline
The phase $ \Phi(p_0,V)$, and its derivatives in equation (\ref{a2}) scale as $1/\alpha$ as the potential becomes broader, $\alpha\to 0$.
The momenta of the initial wave packet (\ref{6}) lie approximately in the range of $\Delta p \sim 1/\Delta x $ around $p_0$,
so that we have $(p-p_0)^n\sim 1/\Delta x ^n$. Distortion-free transmission
will be achieved if the sum in equation (\ref{a2}) can be neglected, i.e., for $\sum_{n=2}^\infty \partial_p^n \Phi(p_0,V)(p-p_0)^n/n! \ll 1$.
This will be the case if we choose
\begin{equation}\label{a3}
\Delta x \sim 1/\alpha ^\gamma, \q 1/2<\gamma < 1.
\end{equation}
 In  equation (\ref{a2}), the remaining term in the square brackets is the classical expression for the distance $\x'$ separating the particle from its freely propagating counterpart
at a given (sufficiently large) time $t$. Indeed, it can be written as [$v_0=p_0/\mu$ and {\crr$v(x,p_0)=q(x,p_0)/\mu$]}
\begin{equation}\label{a4}
\x'\equiv v_0\int_{-\infty} ^\infty\left[\frac{1}{v_0}-\frac{1}{v(x,p_0)}\right]dx,
\end{equation}
 and equation (\ref{1}) reduces to the desired classical result
\begin{equation}\label{a5}
\psi^T(x,t) =
\exp[i \Phi(p_0,V)+ip_0\x']
\psi^0(x-\x',t,p_0). \q
\end{equation}
In a snapshot taken on a large scale  $\sim 1/\alpha$ at a large time $t$, both the transmitted and the freely propagating probability densities,
$|\psi^T(x,t)|^2$ and $|\psi^0(x,t)|^2$, will look like those belonging to point-size particles.
As expected, in the case of a well, $V(x) <0$, $v(x,p) \ge v_0$, and we find the particle lying ahead of
its freely propagating counterpart. Similarly, a barrier, $V(x)>0$, would cause the particle to lag behind the free motion, as is illustrated in Figs. 2{\bf a} and 2{\bf b}. More interesting, however, is the case of  transmission not allowed in classical mechanics, which we will discuss next.
\newline
 \section{Apparently \e{instantaneous} semiclassical tunnelling}
 Next we consider the case of a barrier, $V(x)>0$, and choose all  energies, $E(p) < \text{max}[V(x)]$, to lie  not too close to the barrier top. Now
the local momentum $q(x,p)=\sqrt{p^2-2\m V(x)}$
is imaginary for $x_< < x < x_>$, where $x_{<(>)}$ are the classical turning points, $q(x_{<(>)},p)=0$, and real positive elsewhere.
As $\alpha \to 0$, the semiclassical condition is satisfied everywhere, except in the vicinities of the turning points.
Using the standard connection formulae \cite{Land} it can be shown that, as before,
 \begin{equation}\label{b1}
T(p,V)\approx \exp\left \{i\int_{-\infty}^\infty [q(x,p)-p]dx\right \},
\end{equation}
 except that now $q(x,p)=i|q(x,p)|$ for $x_< \le x \le x_>$, and $|T(p,V)|^2\sim \exp\left (-2\int_{x_<}^{x>}|\sqrt{p^2-2\m V(x)}|dx\right ) \ll 1$,
 so that most of the particles are reflected, and only few are transmitted.
 \newline
Acting as in the previous Section, and assuming that (\ref{a3}) holds,  we find the transmitted wave packet greatly reduced,
narrow compared to the width of the potential,
and somewhat distorted. As before, $\psi^T(x,t)$ is given by equation (\ref{a5}),
but with a {\it complex valued} spatial shift, $\x'=\x'_1+i\x'_2$,
   \begin{align}\label{b2}
\x'_1 &\equiv v_0\int_{-\infty} ^\infty\left\{\frac{1}{v_0}-\R \left [\frac{1}{v(x,p_0)}\right ]\right\}dx,\nonumber\\
\x'_2 &\equiv- v_0\int_{x_<} ^{x_>}\frac{dx}{|v(x,p_0)|} <0. \q\q\q\q\q
\end{align}
{\crr This complex shift has a different effect on the transmitted state.} For a Gaussian wave packet (\ref{-6}) evaluation of the integral (\ref{-2}) yields
\begin{equation}\label{b3}
\psi^T(x,t) = T(p_0,V)\exp[\Delta p^2\x^{'2}_2/4+ip_0\x'_1]\times\psi^0(x-\x'_1,t,p_0+\delta p_0) \q\q\q\q\q
\end{equation}
where $\psi^0(x-\x'_1,t,p_0+\delta p_0)$ is a free wave packet with a mean momentum $p_0+\delta p_0$, and $\delta p_0 = \Delta p^2 \x'_2/2$. Thus, the wave packet is duly delayed by the barrier potential in the classically allowed region.
It also appears to cross the classically forbidden region, which does not contribute to $\x_1'$, \e{instantaneously}.
 Furthermore, traversing
the region increases the particle's mean momentum by $\delta p_0$. This is the well known \e{momentum filtering effect} (see, for example, \cite{Filt}).
Since  higher momenta tunnel more easily, the transmitted particle moves faster than the incident one, and the factor multiplying
$\psi^0(x-\x'_1,t,p_0+\delta p_0)$ in equation (\ref{b3}) is larger than $T(p_0,V)$.
In a snapshot taken at a sufficiently large time $t$ the  (greatly reduced) tunnelling wave packet would
be advanced due to the positive shift induced by the barrier, $\x_1'>0$, as well as owing to the increase in the particle's mean velocity (see Fig. \ref{1}{\bf c}).
 \section{Scattering potential as an \e{interferometer}}
To get an alternative perspective on the three cases shown in Fig. 2,  we change to a different representation by taking a Fourier transform
of the transmission amplitude with respect to $p$, and inserting the result into equation (\ref{-2}). This yields an equivalent expression for the transmitted
state \cite{DSE0},
\begin{align}\label{c1}
\psi^T(x,t)=&\exp[ip_0x -iE(p_0)t]\times \int_{-\infty}^\infty G_0(x-x',t)\eta(p_0,x')dx',
\end{align}
where $G_0(x,t)$ is the freely propagating envelope in equation (\ref{-4a}).
The distribution  $\eta(p_0,x')$ is a sum of a Dirac delta, and a non-singular smooth function,
\begin{equation}\label{c2}
\eta(p_0,x')=\delta(x')+\et(p_0,x'),
\end{equation}
where
\begin{align}\label{c3}
&\et(p_0,x')\equiv  \exp(-ip_0x')\xi(x')=
 (2\pi)^{-1}\exp(-ip_0x')\int_{-\infty}^\infty [T(k,V)-1]\exp(ikx')dk.
\end{align}
With this, the action of a potential $V(x)$ can be understood as follows.
There is a continuum of routes, each labelled by the value of $x'$, via which the particle
can reach its final state. Along each route an envelope $G_0(x,t)$ is enhanced, or
suppressed, by a factor $|\et(p_0,x')|$, acquires an additional phase, $\arg[\et(p_0,x')]$,
and is shifted in space by $x'$. On exit from this \e{interferometer}, all envelopes
are recombined, and  added to $G_0(x,t)$ to produce the transmitted wave
packet in equation (\ref{c1}).  In addition, we have
\begin{equation}\label{c2a}
T(p_0,V) =\int_{-\infty}^\infty\eta(p_0,x')dx',
\end{equation}
so that $\et(p_0,x')$ can be understood to be  the probability amplitude for a particle
with a momentum $p_0$ to experience a spatial shift  $x'$ while crossing a short-range potential  $V(x)$ (see {\crr Supplementary Appendix B}).
\newline
To see how a unique shift appears in the classical limit of Section III, we insert the semiclassical approximation (\ref{a1}) into (\ref{c3}),
and evaluate the integral over $k$ by the method of steepest descent \cite{SD}. This yields
\begin{equation}\label{c4}
\eta(p_0,x') \sim \exp(-ip_0x') \exp[i\Phi(\k(x'),V)+ix' \k(x')]\q\q
\end{equation}
where $\k (x)$ satisfies a condition
\begin{equation}\label{c5}
\partial_k [\Phi(k,V)+k x']|_{k={\k }(x')}=0.
\end{equation}
As a function of $x'$, $\eta(p_0,x')$ has a critical point at $\x'(p_0)$, such that
\begin{equation}\label{c5}
\partial_{x'} [\Phi(\k,V)+\k x'-p_0 x']|_{x'=\x'(p_0)}=\k(\x')-p_0=0,   \q\q
\end{equation}
or, explicitly,
\begin{equation}\label{c6}
\x'(p_0)=-\partial_k \Phi(k,V)| _{k=p_{0}} = \int_{-\infty}^\infty dx
\left[1-\frac{p_0}{q(x,p_0)}\right ].\q\q
\end{equation}
For a particle passing over a well, or above a barrier, $\x'(p_0)$ is real, and we recover the
result (\ref{a4}), but in a new context.
The smooth part of the amplitude distribution, $\et(p_0,x')$ has two important features.
One is a region centred at $\x'(p_0)$ where its oscillations are slowed down, the other is
a finite narrow dip near $x'=0$. The purpose of the dip is to cancel the contribution from the
$\delta$-function in equation (\ref{c2}) (for a similar \e{Zeno peak} occurring  in quantum measurements see \cite{RT}).
The purpose of the stationary region is to replace the now cancelled out $G_0(x,t)$ with $G_0(x-\x',t)$.
Thus, the classical advancement (or delay), $\x'$,
corresponds to a stationary point of the phase of a rapidly oscillating $\eta(p_0,x')$, as shown in Fig. 2{\bf d} and 2{\bf e}. If the incident wave packet is not too
narrow [ $\Delta x > 1/\sqrt{\partial^2_p \Phi(p_0,V))}] $ a single envelope $G_0(x-\x',t)$ is selected
by the integral (\ref{c1}), and the classical picture is restored.
\newline
The case of tunnelling  in Section IV is radically different. The position of $\x'$ is controlled by the particle's mean momentum
$p_0$ and, for $E(p_0) < \text{max} [V(x)]$,  $\x'$ becomes  the position of a complex saddle point off the real $x'$-axis.
On the real  $x'$-axis, the amplitude distribution $\eta(p_0,x')$ rapidly oscillates, and it is impossible to choose
a single real spatial delay, associated with a classically forbidden tunnelling transition (cf. Fig. 2{\bf f}). Destructive interference between the delays
makes the tunnelling amplitude $T(p_0,V)$ exponentially small, yet $\eta(p_0,x')$ is not itself small,  and differs only by a phase
$\exp[-i(p_0-p_0')x']$ from a distribution for a particle with mean momentum $p_0'$ passing over the barrier top  [cf. equation (\ref{c3})] .
This makes semiclassical tunnelling much more delicate, as both the shape and position of the tunnelled wave packet are now determined
by the analytic continuation of the envelope $G_0(x,t)$, $G_0(x-\x'_1-i\x'_2,t)$, into the complex $x'$-plane.
\newline
{\crr In the general case one cannot obtain the transmitted state by a single coordinate shift, be it real or complex valued,
and has to sum the contributions of all the delays by evaluating the integral (\ref{c2a}). In the next Section we discuss a
useful way of doing it.}

\section{ Amplitude distribution of the delays. A pole representation }
One can close the contour of integration in equation (\ref{c3}) in the upper and the lower halves of the complex $k$-plane for
$x'>0$ and $x'<0$, respectively, and obtain $\eta(p_0,x')$ by summing the contributions of pole singularities
of the transmission amplitude.
The poles of $T(k,V)$ fall into two categories \cite{Baz}. Those on the positive imaginary semi-axis
correspond to the bound ($B)$ states, supported by the potential $V(x)$. The poles in the lower half-plane, located symmetrically on both sides
of the negative imaginary semi-axis describe scattering resonances ($R$).  It is sufficient to know all poles positions, $k_n$, and the corresponding residues, $\Res(k_n)$, in order to evaluate all the quantities
of interest. In particular, we have
\begin{align}\label{d1}
\eta(p_0,x') =\delta(x') + i \exp(-ip_0x')
  \times
    \begin{cases}
    \sum_{n_{B}} \Res(k_{n_B}) \exp(ik_{n_B} x'),& x'>0\\
   -\sum_{n_{R}} \Res(k_{n_R}) \exp(ik_{n_R} x'),& x'<0,
    \end{cases}
    \end{align}
\begin{align}\label{d2}
\psi^T(x,t) =\exp[ip_0x -iE(p_0)t]  \{G_0(x,t)&+
 i\sum_{n_{B}} \Res(k_{n_B})\int_0^\infty G_0(x-x',t) \exp[i(k_{n_B}-p_0) x']dx'\nonumber\\
 &-i\sum_{n_{R}} \Res(k_{n_R})\int_{-\infty}^0 G_0(x-x',t) \exp[i(k_{n_R}-p_0) x']dx'\}
    \end{align}
    and
\begin{equation}\label{d3}
T(p,V) =1
- \sum_{n_{B}} \frac{\Res(k_{n_B})}{k_{n_B}-p}
- \sum_{n_{R}} \frac{\Res(k_{n_R})}{k_{n_R}-p}.
\end{equation}
In the above equations the sums are over the
simple
poles corresponding to the bound states ($n_B$), and to the resonances
($n_R$).  The advancement of a classical particle passing over a potential well is already anticipated in equation (\ref{d1}) and (\ref{d2}).
Indeed, an envelope in equation (\ref{c1}) is advanced relative to free propagation, provided we have $x'>0$. Such advanced envelopes would
be present whenever  $V(x)$ is a well, supporting at least one bound state, and would not be there for a barrier where $V(x)>0$ for all $x$.
Special cases where $T(k,V)$ has higher (second) order poles must be treated differently, as discussed in the {\crr Supplementary Appendix C}.
\newline
[Note  that although $\eta(p_0,x')$ is clearly not an analytical function, the asymptotic analysis of the previous Section still applies.
In the case of tunnelling, one could construct the asymptote of  $\eta(p_0,x')$ e.g., on the negative $x'$-axis, analytically continue it into  the entire $x'$-plane, and transform
the contour of integration $\int_{-\infty}^0 dx \to \int_{\Gamma} dx$, making the $\Gamma$ pass through the complex saddle $\x'=\x'_1+i\x'_2$.]
\newline
For an Eckart potential, be it a barrier or a well, the poles' positions and residues are known exactly.
They  occur when the argument of  at least one of the Gamma functions  in the numerator of equation (\ref{-4}) equals  a negative integer, or zero, $-n$, $n=0,1,2,...$. They can, therefore,
 be divided into two groups,
\begin{align}\label{e1}
k^I_n &= i\alpha(s-n), \q n=0,1,2,...,\nonumber\\
k^{II}_n &= -i\alpha(n+s+1), \q \q \q \q
\end{align}
and the corresponding residues are easily found to be given by
\begin{align}\label{e2}
\Res(k^I_n)= i\frac{(-1)^n}{n!}\frac{\alpha \Gamma(-ik^I_n/\alpha+s+1)}{\Gamma(-ik^I_n/\alpha)\Gamma(-ik^I_n/\alpha+1)},\q\nonumber\\
\Res(k^{II}_n)= i\frac{(-1)^n}{n!}\frac{\alpha \Gamma(-ik^{II}_n/\alpha-s)}{\Gamma(-ik^{II}_n/\alpha)\Gamma(-ik^{II}_n/\alpha+1)}.
\end{align}
The cases of a barrier ($U_0>0$) and a well ($U_0<0$) need to be considered separately, as we will do next.
\begin{figure}[h]
\centering
\includegraphics[width=0.6\linewidth]{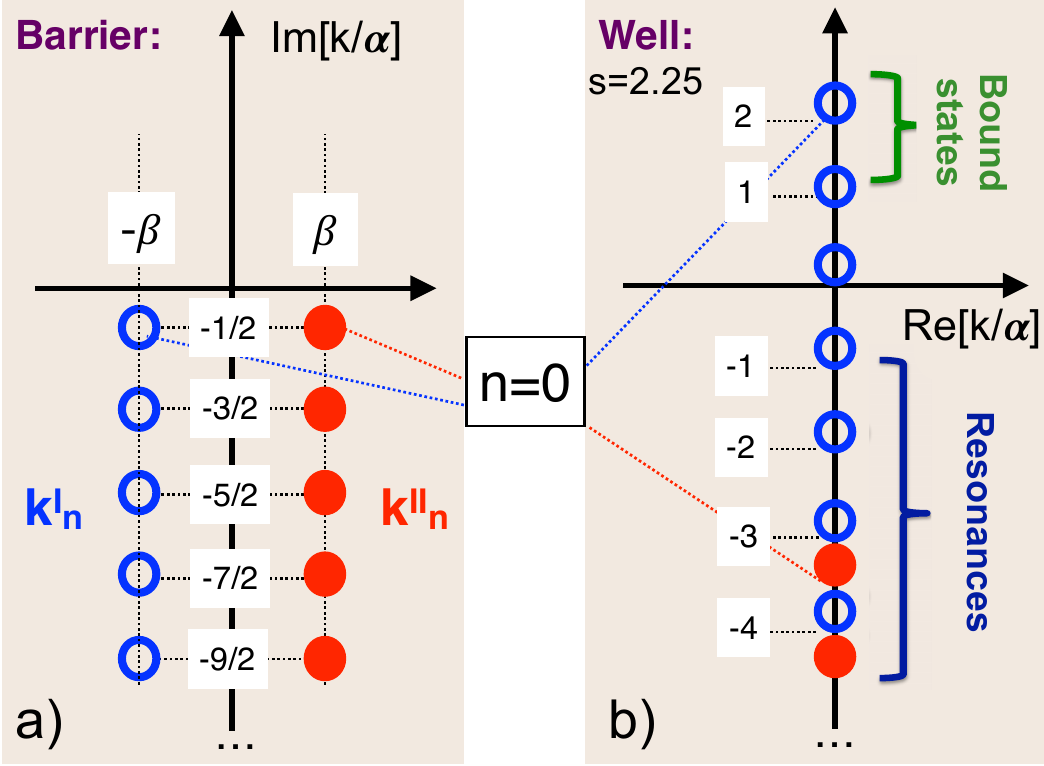}
\caption {{\bf a}) Poles of $T(k,V)$ of the first (open circles) and the second (closed circles) types for an Eckart barrier, $U_0>0$.
{\bf b}) same as {\bf a}), but for an Eckart well, $U_0<0$,  supporting three bound states at  $s=2.25$.
}
\label{2}
\end{figure}
\section{An Eckart barrier}
If $U_0>\alpha^2/8\mu$, we have $s=[-1+i\sqrt{8\m U_0/\alpha^2-1}]/2$, and the poles lie
on two
vertical lines, parallel to the imaginary $k$-axis in the lower half of the complex $k$-plane (see Fig. 3{\bf a}),
\begin{align}\label{e3}
\R[k^{II}_n] &= 2^{-1}\alpha\sqrt{8\m U_0/\alpha^2-1} =-\R[k^{I}_n]\equiv \beta, \nonumber\\
\Ip[k^I_n] &= -\alpha(n+1/2) =\Ip[k^{II}_n], \q n=0,1,2,... \q
\end{align}
It follows from  (\ref{-5a}) that
$\Res[k^I_n]=-\Res^*[k^{II}_n]$,
 and it can be shown (see {\crr Supplementary Appendix D}) that
$\lim_{n\to \infty}\Res[k^I_n]=-\alpha/2\pi$.
 Subtracting the limit, and summing the geometric progression we  find
  \begin{equation}\label{e5}
\tilde \eta(p_0,x\le 0)=
\frac{\exp(-ip_0x)}{2\pi}
\left [F(x)- \frac{ \alpha\sin(\beta x)}{\sinh(\alpha x/2))}\right ],\q
 \end{equation}
 where $F(x)$ is given by a convergent series,
   \begin{align}\label{e6}
F(x)&\equiv 4\pi\sum_{n=0}^{\infty}\exp[(n+1/2)\alpha x]\times \Ip\{[\Res(k^I_n)+\alpha /2\pi]\exp(-i\beta x)\}.
 \end{align}
\newline
At  $U_0=\alpha^2/8\mu$,  $s=-1/2$, and the poles coalesce on the negative imaginary semi-axis
\begin{align}\label{e3}
\R[k^I_n] &=\R[k^{II}_n]=0, \nonumber\\
 \Ip[k^I_n] &=\Ip[k^{II}_n]= -\alpha(n+1/2).
 \end{align}
For low barriers,  $0<U_0<\alpha^2/8\m$, $s$ remains real and negative.  Thus, as $U_0$ increases, the poles of the first kind,  ($I$),
move up  the negative imaginary $k$-axis, while the poles of the second kind,  ($II$), move down the same axis.
As $U_0\to 0$, we find $k^I_0 \to 0$, and  $k^{II}_n\to -i\alpha n=k^I_{n+1}$.
The $k^I_0 \to 0$ pole corresponds to the first {\it virtual state}, ready to become the first bound state when the potential
becomes a shallow well. Virtual states of this type were described as \e{long-lived} in \cite{Baz}, and we will return to them
in Section XII.
\section{An Eckart well}
As $U_0$ decreases further and becomes negative,  $V(x)$ forms a one dimensional well, which always supports at least one bound state \cite{Land}.
The poles of the first kind continue their upward motion along the imaginary axis, and for $s>M$, $M=0,1,2,...$, $M+1$ of them
lie on the positive semi-axis, where they correspond to the $M+1$ bound states, $k_{n_B}=k^I_n$, $n_B=s-n$, $n=0,1,...,M$.
The poles of the second kind,  $k^{II}_n = -i\alpha (n+s+1)$ move down the imaginary axis (cf. Fig. 3{\bf b}). We note that some of the poles
on the negative semi-axis coalesce whenever $s$ takes an integer, or a semi-integer value.
\newline
 The analysis is the simplest in the case where $s$ takes an integer value, $s=M$,
[$U_0=-\alpha^2M(M+1)/2\m$],
$M=1,2,3,...$,
and a new bound state is about to enter into the well. The only singularities are the $M$ bound states poles, $k^I_{n}$, $n=0,...,M-1$, and $T(k,V)$ remains finite elsewhere, since the singularities of the numerator in equation (\ref{-4}) are cancelled,
by the two Gamma functions in the denominator.
Thus,
the series (\ref{d1})-(\ref{d3}) become finite sums, and we have
\begin{align}\label{ff1}
\eta(p_0,x') =\delta(x') + i \exp(-ip_0x')
  \times
    \begin{cases}
    \sum_{n=0}^M \Res(k^I_n) \exp(ik^I_n x'),& x'\ge 0\\
   \q 0,& x'<0,
    \end{cases}
\end{align}
where $k_n^I=i\alpha (M-n)$, and
\begin{equation}\label{ff2}
\Res(k^I_n)=(-1)^n i\alpha \frac{(2M-n)!}{n!(M-n)!(M-n-1)!}.
\end{equation}
Note that at these integer vales of $s$  the well is transparent for all incident momenta, $|T(p,V)|=1$.
\newline
For $s=M+1/2$, $M=0,1,...$ there is a sequence of double poles, whose residues are given in {\crr Supplementary Appendix C}.
Note that the coalescence of the poles does not produce any visible feature in the behaviour of either $\eta(p_0,x')$ or $T(p,V)$.
{\color{black}
\section{Is there an effective  range of spatial delays?}
There is clearly no unique spatial delay describing scattering beyond the classical limit.
But is there a characteristic {\it range} of delays one can use, e.g., for a qualitative description of quantum tunnelling?
Consider  first a probability distribution with a well defined  range, e.g.,
$\rho(x) = \exp(-\gamma x)$ for $x\ge 0$, and $0$ otherwise. It vanishes for $x\gg 1/\gamma$,
and obviously has a \e{size}  $\sim 1/\gamma$. This can be checked by evaluating its moments,
 $\overline{ x^m} =\int x^m\rho(x) dx/\int \rho(x) dx$, $m=1,2,...$, and noting that $\overline{ x}=1/\gamma$ and
$\sigma \equiv \sqrt {\overline{ x^2}-\overline{ x}^2}=\sqrt{2}/\gamma$, yield estimates for  the position and the width of the  region
which contains most of the distribution.
\newline
This is no longer true if a distribution is allowed to change sign (for a more detailed discussion see \cite{negat}).
Consider next a different \e{distribution},
 \begin{equation}\label{f1a}
\rho(x)=(1/2+\epsilon)\exp(-x)-\exp(-2x), \q x\ge 0,
\end{equation}
whose size, defined as above, should be of order of $1$, since the slowest decaying term in the sum (\ref{f1a}) is $\exp(-x)$.
An integral $I(z)\equiv \int_{0}^\infty\exp(-x^2/z^2)\rho(x) dx$,
now restricted to an effective range of the order of $z$,
should converge to $I=\int_0^\infty  \rho(x) dx =\epsilon$ as $z\to \infty$.
The question is how fast?
 Expanding $\exp(-x^2/z^2)\approx 1-x^2/z^2$ and requiring that the contribution of the second term
 to
be negligible,  yields an estimate
$z^2 \gg {\left |{\overline {x ^2}}\right |}$.
 Thus, for $\epsilon \to 0$ we have $z \gg 1/\sqrt \epsilon\to \infty$,  at odds
 with the expected condition $z \gg 1$.
Clearly, a much longer integration range is required for recovering a very small result to a good relative accuracy.
\newline
A similar problem would occur in trying to estimate how many delays must be taken into account in order to accurately
reproduce the transmission amplitude $T(p,V)$ in the case of tunnelling.
Like $\rho (x)$ in equation (\ref{f1a}), the distribution $\xi(x')$
 in equation (\ref{c3}) is a sum of exponential terms  [cf. equation (\ref{d1})], and for an Eckart barrier with $U_0 >\alpha^2/8\mu$
the $\exp(ik^I_0x')$ and $\exp(ik^{II}_0x')$ have the slowest decay rate of $\Ip[k^I_0] =\alpha/2$.
Can it then be said that the transmission amplitude  is a result of interference between delays in a range $-\alpha \lesssim x' <0$?
The previous example points toward a problem which may arise, especially where a small result is obtained
due to cancellations between terms which are not small, as happened in the case of tunnelling (see Fig. \ref{3}{\bf b}).
Acting as in the above, we can define complex valued moments of the distribution $\eta(p_0,x`)$,
\begin{equation}\label{f1}
\overline{x'^{m}}\equiv \int x'^m \eta(p_0,x')dx'/\int \eta(p_0,x')dx'=
T(p_0,V)^{-1}(i\partial_p)^m T(p_0,V),\q m=1,2,...
\end{equation}
and obtain a necessary (but not sufficient) condition to have $|T(p_0,V|z)-T(p_0,V)|/|T(p_0,V)| \ll 1$,
\newline
$T(p_0,V|z)\equiv \int_{-\infty}^0 \exp(-x^{'2}/z^2)\eta(p_0,x')dx'$,
\begin{equation}\label{f2}
z \gg \sqrt{|\overline {x'^2}|} = \sqrt{|\partial^2_pT(p_0,V)/T(p_0,V)|}\equiv R(p_0,V),\q\q
\end{equation}
For semiclassical tunnelling one finds $z \gg  \sqrt{\x_1^2+\x_2^2}$, so the {range} $R(p_0,V)$,  defined in this manner,  is of order of the modulus
of the complex delay in equation (\ref{b2}) , $|\x|$.
\newline
We note that the dependence of $R(p_0,V)$ on $p_0$ clearly frustrates our attempts
to define an effective range of integration in equation (\ref{c2a}), based on estimating the \e{size} of $\xi(x')$ in equation (\ref{c3}),
i.e., by taking into account only the properties of the potential, and ignoring the value of the particle's momentum.
We will return to this subject in Section XII.
\section{Spatial delays vs. detection times}
One way to quantify the effect produced by a potential on a transmitted particle is to compare, at a given time $t$,  the positions of the centre of mass (COM)
of the transmitted density with that of a freely propagating state. For the COM delay we have
 \begin{equation}\label{y1a}
\delta x_{COM}(t)\equiv x^{T}_{COM}(t)-x^{0}_{COM}(t),\q
\end{equation}
where
 \begin{equation}\label{y1}
x^{T,0}_{COM}(t)=\int x|\psi^{T,0}(x,t)|^2 dx/\int |\psi^{T,0}(x,t)|^2 dx, \q
\end{equation}
With the help of  equation (\ref{a5}) we readily obtain the classical result of Section III,
$\delta x_{COM}(t) = \x'$.
For semiclassical tunnelling of Section IV, from equation (\ref{b3})  we find
 \begin{align}\label{y3}
\delta x_{COM}(t)=&
[\x'_1+(v_0+\delta v_0)t +x_0] -[v_0t +x_0]
= \x'_1+\delta v_0t,
\end{align}
where $\delta v_0=\delta p_0/\mu$ is the increase in the velocity due to the momentum filtering.
In the general case the increase in the transmitted particle's mean velocity can be evaluated as
\begin{equation}\label{y4a}
\delta v_0= \delta p_0/\mu=
\frac{\int (p-p_0)|T(p,V)|^2|A(p-p_0)|^2dp}{\mu \int |T(p,V)|^2|A(p-p_0)|^2dp}. \q
\end{equation}
For a large enough $t$, the term $\delta v_0t$ will dominate the r.h.s. of equation (\ref{y3}). However, $\x_1$  can still be determined,
by comparing the position of the COM of $\psi^T(x,t)$ with that of a free particle with a higher initial momentum,
$\psi^0(x,t,p_0+\delta p_0)$, and no additional spatial shift. Note that in equation (\ref{y3}) one would expect $v_0$ to be multiplied by the time interval between $t$ and the moment
the tunnelling particle enters the classical allowed region $x>x_>$ to the right of the barrier. Yet, $t$ is the {\it entire} time of motion, as is illustrated
in Fig. \ref{1}{\bf c}.
\newline
So far we discussed the spatial shifts, since in equation (\ref{c3}) we relied on the Fourier transform of the transmission
amplitude $T(p,V)$ with respect to the momentum $p$.
Of course, knowing the positions of the COM's at a time, as well as their velocities,
it is easy to  compare the times $\tau(p_0,V)$ and $\tau(p_0,V=0)$
at which the majority of the particles would arrive at a fixed detector with and without the potential in place.
In the classically allowed case we have
\begin{equation}\label{y4}
 \tau(p_0,V)-\tau(p_0,V=0) = -\x'/v_0.
\end{equation}
In the case of tunnelling,  a comparison with free motion at  $p_0+\delta p_0$,  yields
\begin{equation}\label{y5}
\tau(p_0,V)-\tau(p_0+\delta p_0,V=0) \equiv -\R[\x']/v_0.
\end{equation}
Both equation (\ref{y4}) and (\ref{y5}) refer to time intervals, which can in principle be measured, yet their interpretation
is different. The l.h.s. of equation (\ref{y4}) can be understood  as the \e{delay experienced by a classical particle in the potential},
since the particle's  position can be determined throughout the transmission sufficiently accurately, and without disturbing
the transition. Such an interpretation is not available for equation (\ref{y5}), where the transmitted state is seen to be
{\it reshaped} via interference mechanism of the previous Section  (for more details see \cite{DSE0}).
 The same is true for any transition resulting from the interference between various delays induced by the potential.
 If so, the measured time of arrival at a detector cannot reveal the delay induced by the barrier.
Its value {\it must} remain indeterminate \cite{DSE1} in accordance with the Uncertainty Principle \cite{FeynL},
 just like the slit chosen by a particle in a Young's double slit experiment.
\section{The \e{phase time}}
Whether surprisingly, or not so surprisingly, it is the $\overline{x'}$ in equation (\ref{f1}) which can be measured in an experiment.
By increasing the wave packet's coordinate width
$\Delta x$ [cf. equation (\ref{-6}) and 
{\crr (S$1$) in Supplementary Appendix A]},  one can prepare a wave packet,  broad in the coordinate space, and narrow
in the momentum representation.
{\color{black}
Expanding
the broad envelope in a Taylor series, $G_0(x-x',t) \approx G_0(x,t) -\partial_xG_0(x,t)x'$
yields \cite{DSE0}
 \begin{align}\label{f3}
\delta x_{COM}&(t)
\approx \R[\overline{x'}] +
 2 \Ip[\overline{x'}]\Ip\left [\int x G_0^*(x,t)\partial_x G_0(x,t) dx\right ].
\end{align}
The last term in equation (\ref{f3}) is recognised as the distance gained  due to the momentum
filtering, discussed in the previous section, and equation (\ref{f3}) can be rewritten
 \begin{equation}\label{f3a}
\delta x_{COM}(t)\approx v_0 \tau_{phase}+\delta v_0 t,\q\q\q
\end{equation}
where $\tau_{phase}\equiv v_0^{-1}\partial_p \Phi(p_0,V)$ [$T(p,V)=|T(p,V)|\exp[i\Phi(p,V)]$] is known
as the \e{phase time} (see, e.g., \cite{PHASE}), and $\delta v_0$ is obtained
by taking the limit $\Delta p\to 0$ ($\Delta x \to \infty$) in equation (\ref{y4a}),
 \begin{align}\label{f3b}
\delta v_0 \approx 2\partial_p \ln|T(p_0,V)|&\frac{\int (p-p_0)^2|A(p-p_0)|^2dp}{\mu \int |A(p-p_0)|^2dp}
=\Ip[\overline{x'}] \Delta p^2/2\mu.
\end{align}
equation (\ref{f3a}) is valid for any potential, as long as the incident wave packet is broad enough in the coordinate space.
Thus, in the classical limit we recover equation (\ref{a4}), and $\tau_{phase}$ becomes the excess
time, positive or negative, spent by the particle's trajectory in the potential.

{\cbb For semiclassical tunnelling $\R[\overline{x'}]=\x_1$  is given by the first of equations (\ref{b2}), so there is no contribution 
to $\tau_{phase}$  from the classically forbidden region, $x_< < x <x_>$.
With the time to cross the forbidden region apparently shorter than the time it takes to cross it at the speed of light,
 one seems to have a dilemma. Either Einstein's relativity has no say over what happens in classically forbidden quantum transitions (see, e.g., \cite{Nimtz}),
or there must be a good reason why one should refrain from deducing the duration spent in the barrier from the distance $x^T_{COM}(t)- x^0_{COM}(t)$. Maybe the problem is with the non-relativistic Schr\"odinger equation used in the calculation?
No, the use of the fully relativistic Klein-Gordon \cite{{Deutsch},{DSrel}} and Dirac \cite{Dumont2} showed the same 
\e{apparently superluminal} advancement of the (greatly reduced) transmitted wave packet. There is some consensus that the \e{superluminal} transmission results form reshaping of the initial wave packet, whereby the 
transmitted particles come from its front tail \cite{{Diener}, {Petersen1},{Deutsch},{DSrel},{DSSh}} and causality is never violated.
(The authors of  \cite{Dumont2} reject the former suggestion, but agree in that no \e{superluminal signalling} is possible.)
\newline
As far as we can see the problem with the \e{phase time} is as follows.
With many envelopes in equation (\ref{c1}) interfering destructively, one cannot determine a unique spatial shift induced by
the barrier \e{interferometer}. In all routes across the barrier potential $V(x)>0$ none of the  envelopes $G(x-x',t)$ in equation (\ref{c1}) are {\it advanced} even relative to the non-relativistic free motion. The average $\overline{x'}$ obtained with an alternating distribution in equation (\ref{f1}) cannot be used in the same way  as the unique classical value of Sect. III. One may recall the double slit conundrum.
The interference picture is clearly observable, yet there is no way of telling which of two the holes has been used by the particle
(See also Supplementary Appendix E).}
\section{Scattering by a shallow Eckart well}
The pole representation (\ref{d1}) turns out to be impractical for a deep semiclassical well, supporting many bound states.
The magnitudes of the residues in equation (\ref{e2}) become prohibitively large (see Fig. 4{\bf a}), and a large amount of cancellation
is required to produce the classical result (\ref{a5}).
\begin{figure}[h]
\centering
\includegraphics[width=0.5\linewidth]{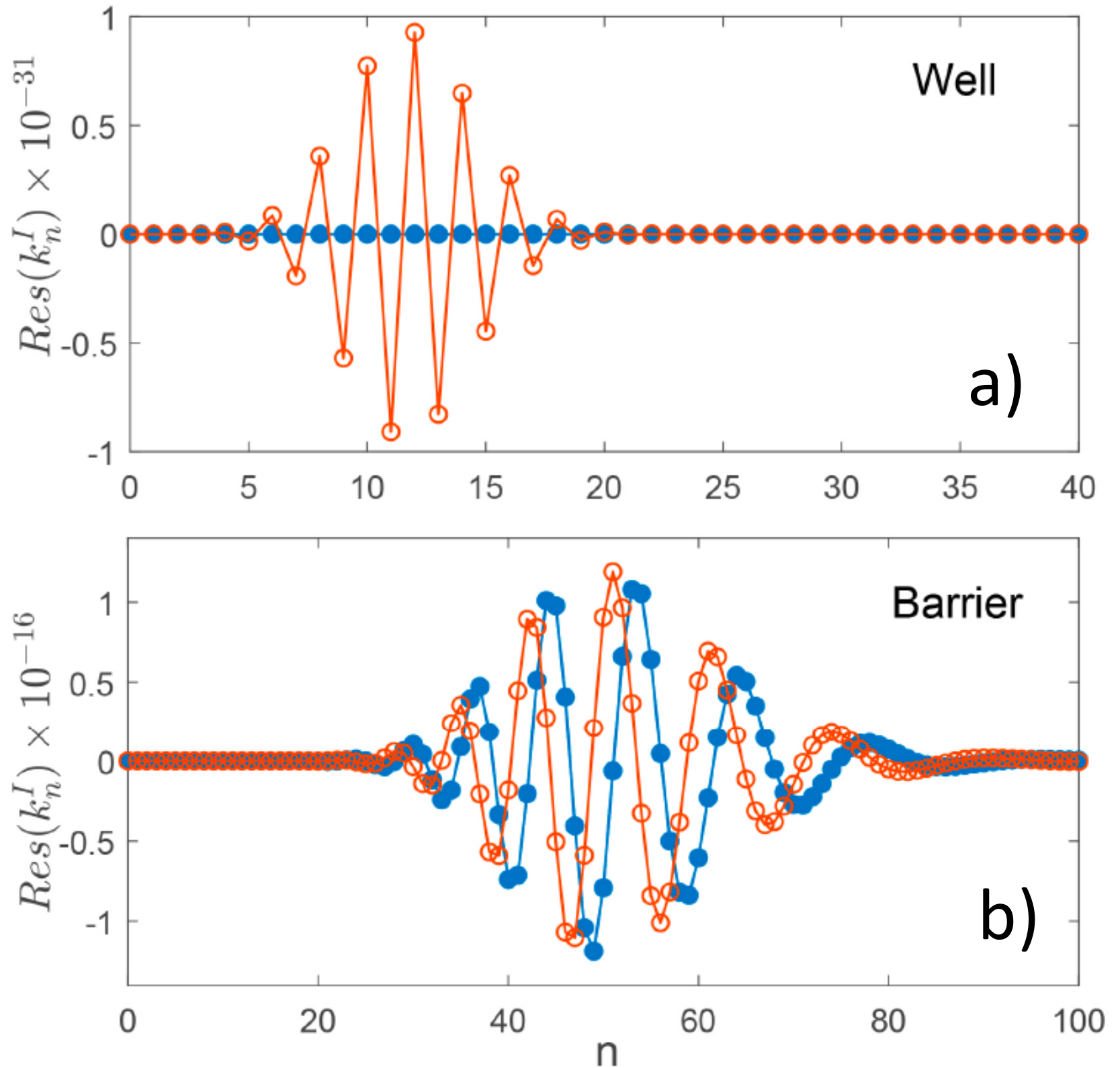}
\caption {{\bf a}) Real (closed circles) and imaginary (open circles) part of the residues
for an Eckart well, supporting $42$ bound states, $\uu=-861$, $s=41$.
Note that here are no singularities
in the lower half-plane.
Note also
the scale on the vertical axis.
{\bf b}) Same as a) but for a barrier with $\uu=2485$, $s\approx -0.5+70,5i$.
The poles of $T(k,V)$ are distributed as in \textbf{Figure 3a}, and there are no singularities
in the upper half of the $k$-plane.
}
\label{3}
\end{figure}
The representation is, however, useful in the case of a shallow well,
where the semiclassical approximation (\ref{a5}) cannot be applied.
\newline
For a Gaussian wave packet (\ref{-3}) the integrals in equation (\ref{d2}) can be expressed in terms of the error function (see {\crr Supplementary Appendix A}),
and for $s=M$ we have
 \begin{align}\label{g1}
\psi^T(x,t)& =\exp[ip_0x -iE(p_0)t]\times
  \{G_0(x,t)+
 \sum_{n=0}^{M-1} \Res(k_{n^I})\large{\textfrak{G}^B(x,k^I_n,p_0)}\}\n
    \end{align}
where $\large{\textfrak{G}(x,k^I_n,p_0)}$ is given by {\crr equation 
(S$3$) in Supplementary Appendix A}.
We note that equation (\ref{g1}) } is exact, and holds for all Gaussian initial states.
An example, where a broad Gaussian wave packet ($\Delta x \gg  1/\alpha$)  crosses an Eckart well, supporting $5$ bound sates is shown in Fig. 5.
\begin{figure}[h]
\centering
\includegraphics[width=0.5\linewidth]{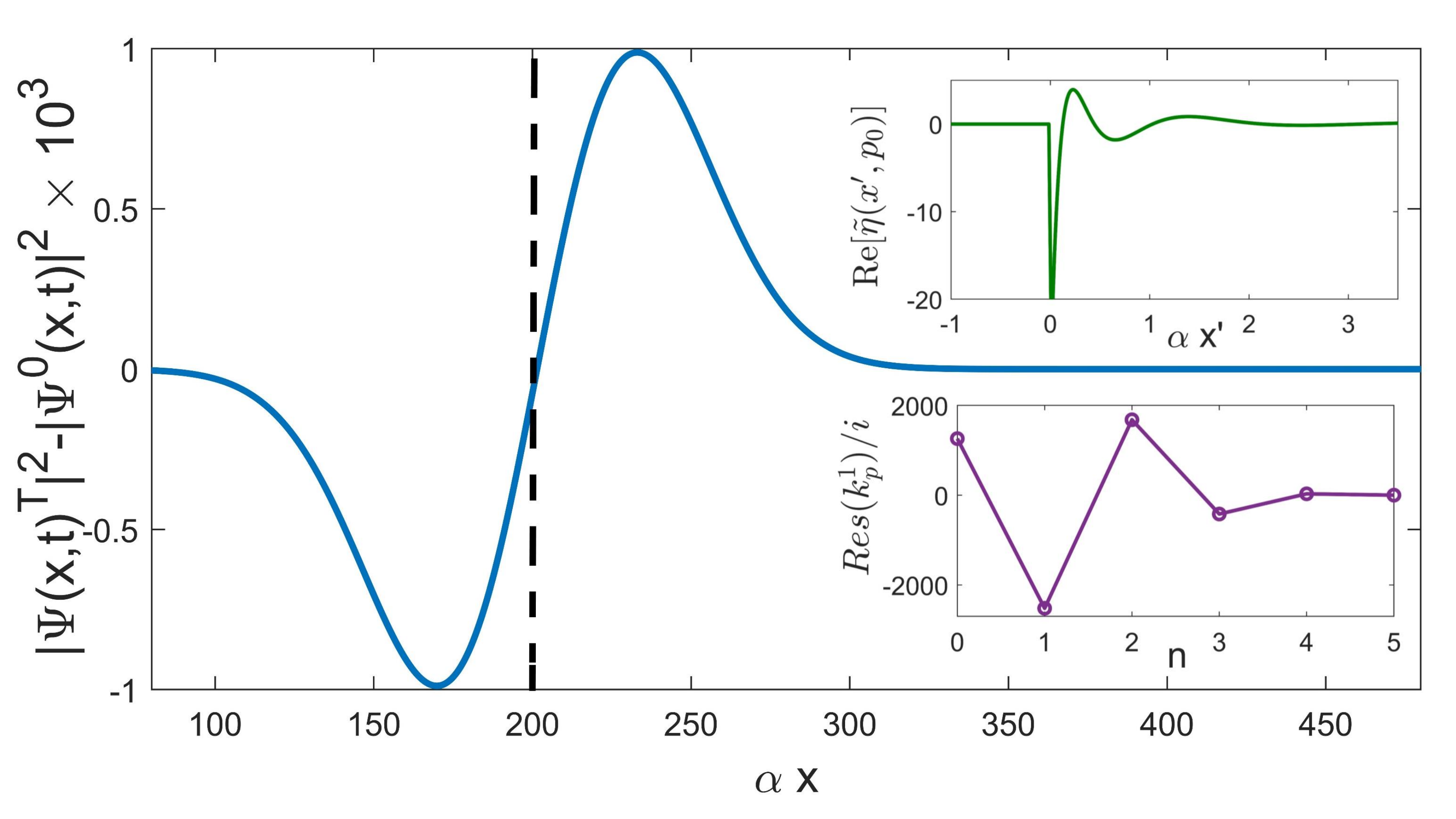}
\caption {The difference between the probability densities of a free wave packet,  $\pp_0=0.5$, $\Delta \pp=0.1$, $\xx_0=-100$, $\tt=650$ and one transmitted across a shallow Eckart well with $s=5$ $(\uu=-15)$.
The COM of the transmitted wave packet is advanced by $\delta x_{COM}(t) = 4.0768$ relative to the COM of a freely propagating one (vertical dashed).
Also shown in the insets are the real part of $\overline{\eta}(p_0,x')$ in equation (\ref{d1}), and the residues of the five poles, which contribute
to equation (\ref{ff1}).
}
\label{4}
\end{figure}
\newline
This analysis can be extended to wells with $s$ close to an integer $M$, $|s-M|\ll 1$.
Below we consider the case $s\sim 1$, where the second bound state is about to enter the well.
(The cases $M\sim 2,3,...$ can be analysed in a similar manner, see {\crr Supplementary Appendix F.})
It is sufficient to include the contributions from the first two poles
of the first kind, $k^I_0 \approx i \alpha s$ and  $k^I_1 \approx i \alpha (s-1)$. Thus, for the delay distribution from equation (\ref{d1})  we have
 \begin{align}\label{g2}
\eta(p,x')&\approx \delta(x')-
2\al\exp[-(\al s+ip)x']\theta(x')
+ \al (s-1)\exp\{-[\al(s-1)+ip]x'\}
\times [\theta(x')\theta(s-1)+\theta(-x')\theta(1-s)],\q\q\q
    \end{align}
 where $\theta(x)=1$ for $x\ge 0$ and $0$ otherwise.
Integrating equation (\ref{g2}) for $ |s-1|,p/\al \ll 1$, yields
 \begin{equation}\label{g3}
T(p,V)\approx -\frac{ip/\al}{(s-1)+ip/\al}+2(s+ip/\al-1)
\end{equation}
The first ( Breit-Wigner) term ensures that, for a fixed $p$, $|T(p,V)|$ is peaked around
$s=1$ with a width $2p/\al$. The second term needs to be taken into account when calculating the derivatives with respect to $p$ at $s=1$.
In particular, for the COM delay of a slow broad wave packet from (\ref{f3}) we obtain,
\begin{equation}\label{g4}
\delta x_{COM}
- \delta v_0 t
\approx \frac{\al(s-1)}{\al^2(s-1)^2+p_0^2}+\frac{2}{\al},\q
\end{equation}
where $ \delta v_0$ given by equation (\ref{f3b}) is the increase in the transmitted particle's mean velocity due to momentum filtering, already
discussed in Sections IV and X.
Thus,  a broad wave packet with $p_0\ll \al$ is advanced relative to free propagation at $p_0+\delta p_0$ by about
the well's width, $\sim 1/\al$.
However, (see Fig. \ref{5a}),  the largest {advancement}, $\sim 1/2p_0 \gg 1/\al$,
is achieved for $s \approx 1+p_0/\al$, where there exists a shallow
level with an energy $E\approx -p_0^2/2\mu$, and $\eta(p_0,x')$ has a long tail extending into the $x'>0$ region.
Similarly, the largest {delay}, $\sim 1/2p_0$ occurs for $s \approx 1-p_0/\al$,
where there exists a {\it virtual state} \cite{Baz}, and  $\eta(p_0,x')$ extends far into the $x'<0$ region.
Finally, for $|s-1|$, $p_0/\al\ll 1$ and $|s-1|\gg  p_0/\al$ we have $
\delta x_{COM}
- \delta v_0 t
\approx 1/\al(s-1)$,
in agreement with the slowly decaying exponential term $\exp[-\al(s-1)]$ in equation (\ref{g2}).
\newline
One can also try to estimate  the effective range the delays as proposed in Section IX.
With the help of equations (\ref{f1}) and (\ref{g2}) one finds
\begin{equation}\label{g5}
\overline {x'^2} \approx \al^{-2} T^{-1}(p_0,V)\left [\frac{-4}{(s +ip_0/\al)^3}+\frac{2(s-1)}{(s-1 +ip_0/\al)^3}\right ].\q\q
\end{equation}
\begin{figure}[h]
\centering
\includegraphics[width=0.5\linewidth]{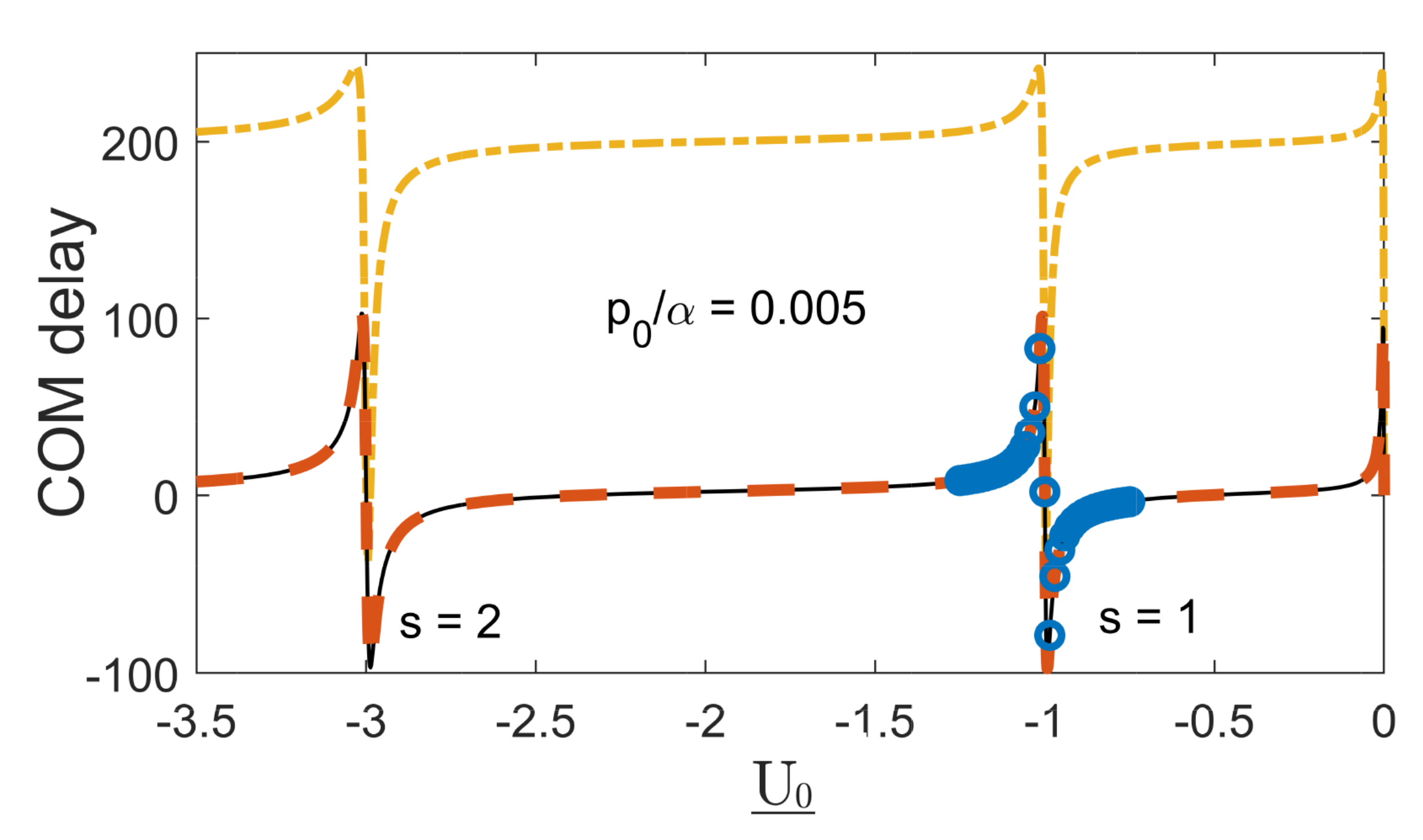}
\caption {Centre-of-mass delay of a wave packet with $\pp_0=0.005$, $\Delta \pp=0.001$, $\xx_0=-3*10^3$,  at
$\tt=2*10^6$ vs. the well's depth $\uu >0$,  calculated using equation (\ref{y1a})-(\ref{y1}) (solid), equation (\ref{f3}) (dashed), and equation (\ref{g4}) (dot-dashed)
}
\label{5a}
\end{figure}
For $s=1$ and $p_0/\al \ll 1$, the range $R(p_0,V)$ in equation (\ref{f2}) equals twice the wells width, $R(p_0,V)=\sqrt{|\overline {x'^2}|} \approx 2/\al$.
This is to be expected, since there $\eta(p_o,x')$ in equation (\ref{g2}) consists of a single exponential with a decay rate $\alpha$.
The range peaks at $|s-1|=p_0/\al$, reaching there the largest value $R(p_0,V) \approx 1/p_0$ as shown in Fig. \ref{5}.
For $|s-1|\gg  p_0/\al$ one may expect the range to be given by the largest decay rate in equation (\ref{g2}), i.e.,
$R(p_0,V) \sim 1/\al |s-1|$. The correct answer is, however, $R(p_0,V) \sim 1/\al \sqrt{|s-1|}$ (see Fig. \ref{5}),
since the value of $|T(p_0,V|$, determined by all three terms in equation (\ref{g2}), itself falls off as $1/|s-1|$.
This, as we have already seen in SectionVII, is a common problem with estimating the effective range of an  alternating distribution.
An integration range, defined in this manner, depends not only on the apparent \e{size} of integrand
(in this case, $1/|s-1|$) but also on the value of the integral, which can itself be small due to cancellations.
\begin{figure}[h]
\centering
\includegraphics[width=0.5\linewidth]{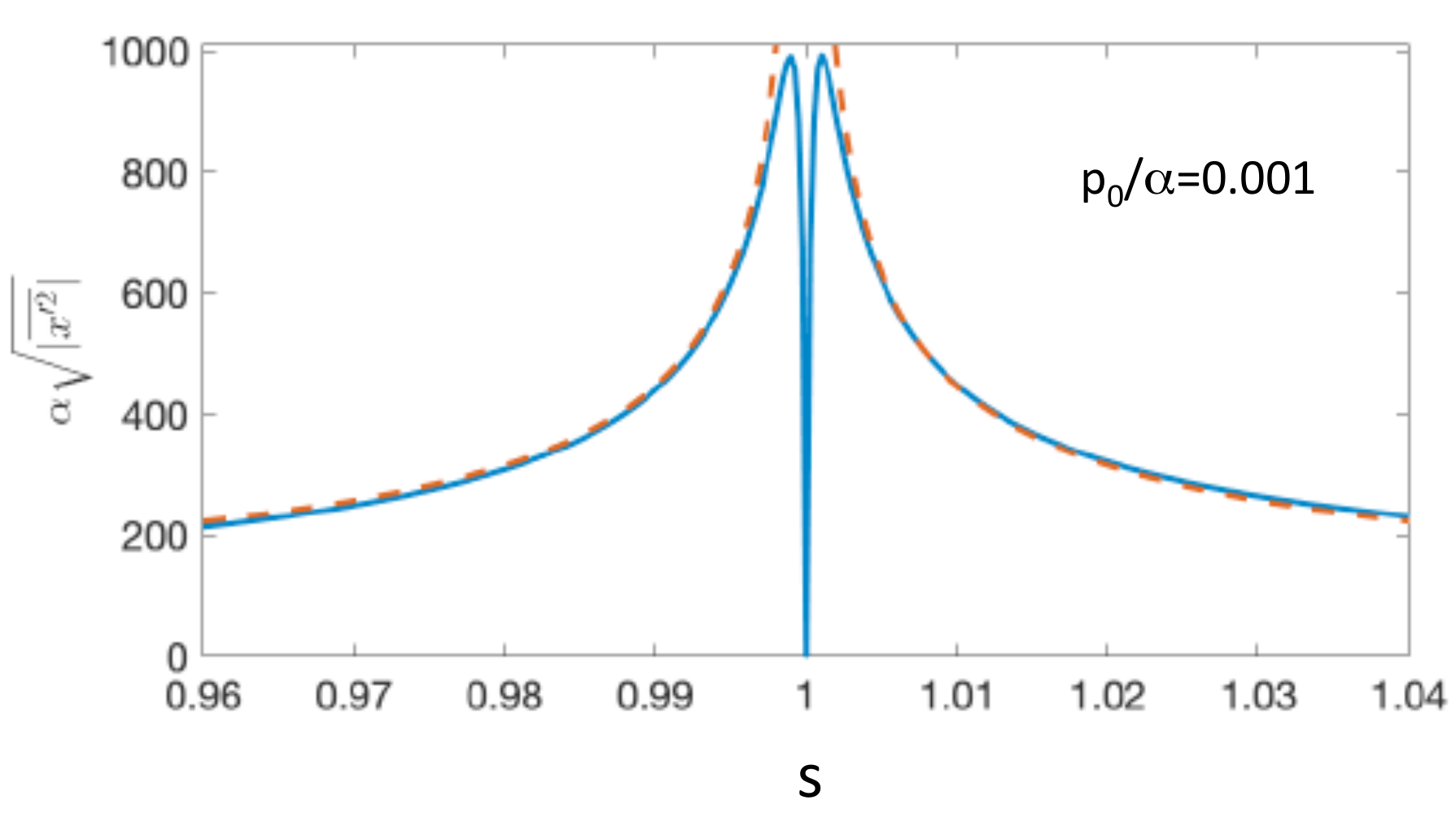}
\caption {Effective range of integration in equation (\ref{c2a}), $R(p_0,V)=\sqrt{|\overline {x'^2}|} \approx 2/\al$
vs. $s$ in equation (\ref{-2}), for a shallow Eckart well, and $p_0/\al =0.001$ (solid).
Also shown by the dashed line is an approximation $R(p,V) \approx \sqrt{2/(p_0|s-1|)}$.
}
\label{5}
\end{figure}
\section{Scattering by a low Eckart barrier}
In the case of  a high semiclassical barrier, the pole representation (\ref{d1}) has a similar problem (see Fig. 4{\bf b}).
For a low barrier, however, the pole expansion is more useful, as illustrated in Fig. \ref{6}.
\begin{figure}[h]
\centering
\includegraphics[width = 0.5\textwidth]{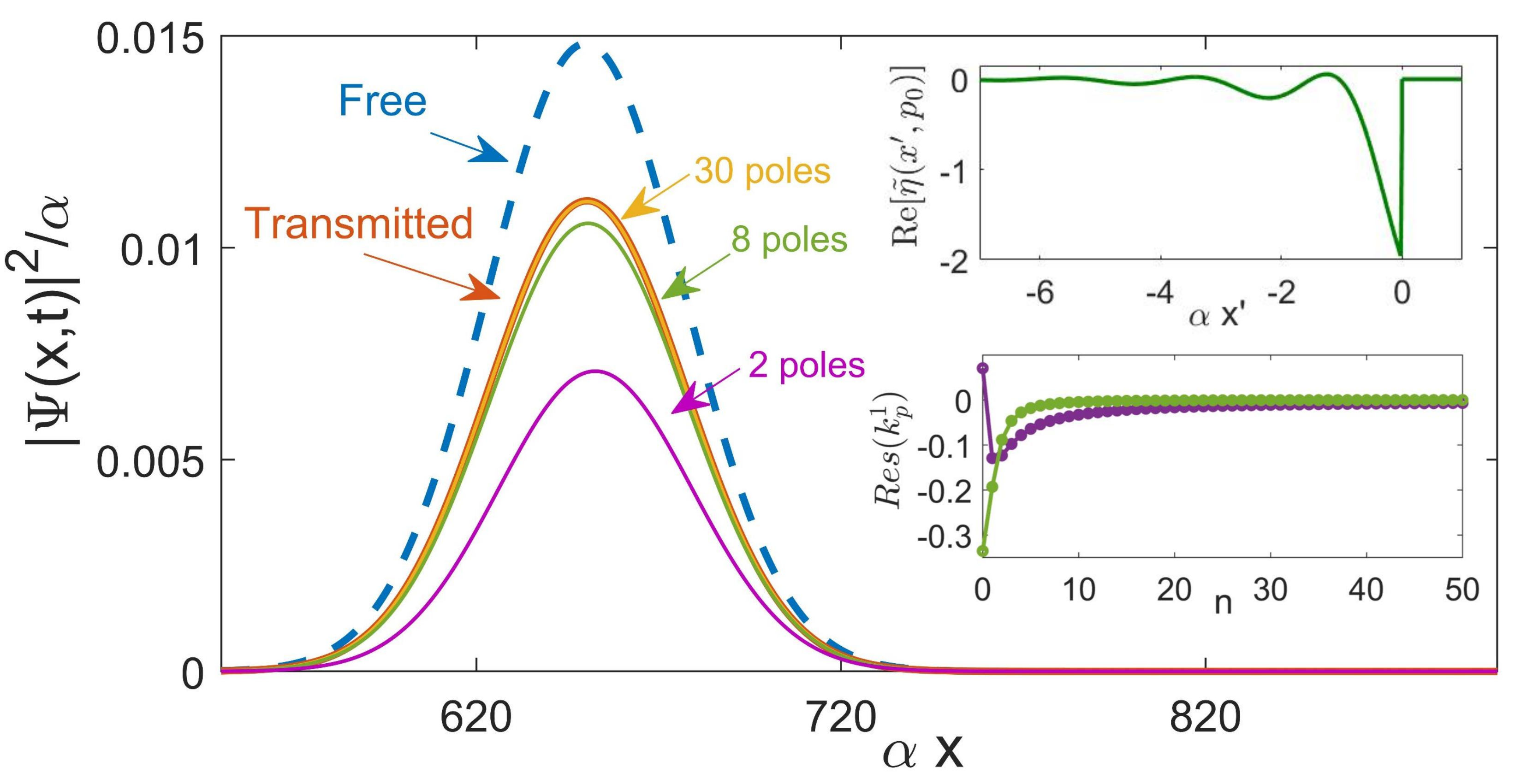}
\caption { A wave packet with $\pp_0=1.5$, $\Delta \pp=0.1$, $\xx_0=-100$,  at
$\tt=500$, transmitted across a low barrier, $\uu=1$ $(s=-0.5+1.323i)$.
Also shown are the results of calculating $\psi^T(x,t)$ in equation (\ref{d2}) using $2$, $8$, and $30$ poles,
and the free wave packet (dashed). The insets show the real part of $\overline{\eta}(p_0,x')$ in equation (\ref{d1}), and the residues of the first $50$ poles.}
\label{6}
\end{figure}
Low energy scattering by a low barrier can be analysed by the method of the previous Section.
For $-1/2< s < 0$, the poles remain on the negative imaginary axis,  and for $|s|\ll 1$ the  delay is dominated
by the presence of the last virtual state, which joins the continuum when the well ceases to exist.
Accordingly, we have
 \begin{equation}\label{h1}
\eta(p_0,x')\approx \delta(x')-
 \al s\exp[-(\al s+ip_0)x']\theta(-x'),\q
    \end{equation}
 \begin{equation}\label{h2}
T(p,V)\approx \frac{ip/\al}{s+ip/\al},
\end{equation}
and from (\ref{f3})
\begin{equation}\label{h3}
\delta x_{COM}
- \delta v_0 t
\approx \frac{\al s}{\al^2s^2+p_0^2}.
\end{equation}
\begin{figure}[h]
\centering
\includegraphics[width = 0.5\textwidth]{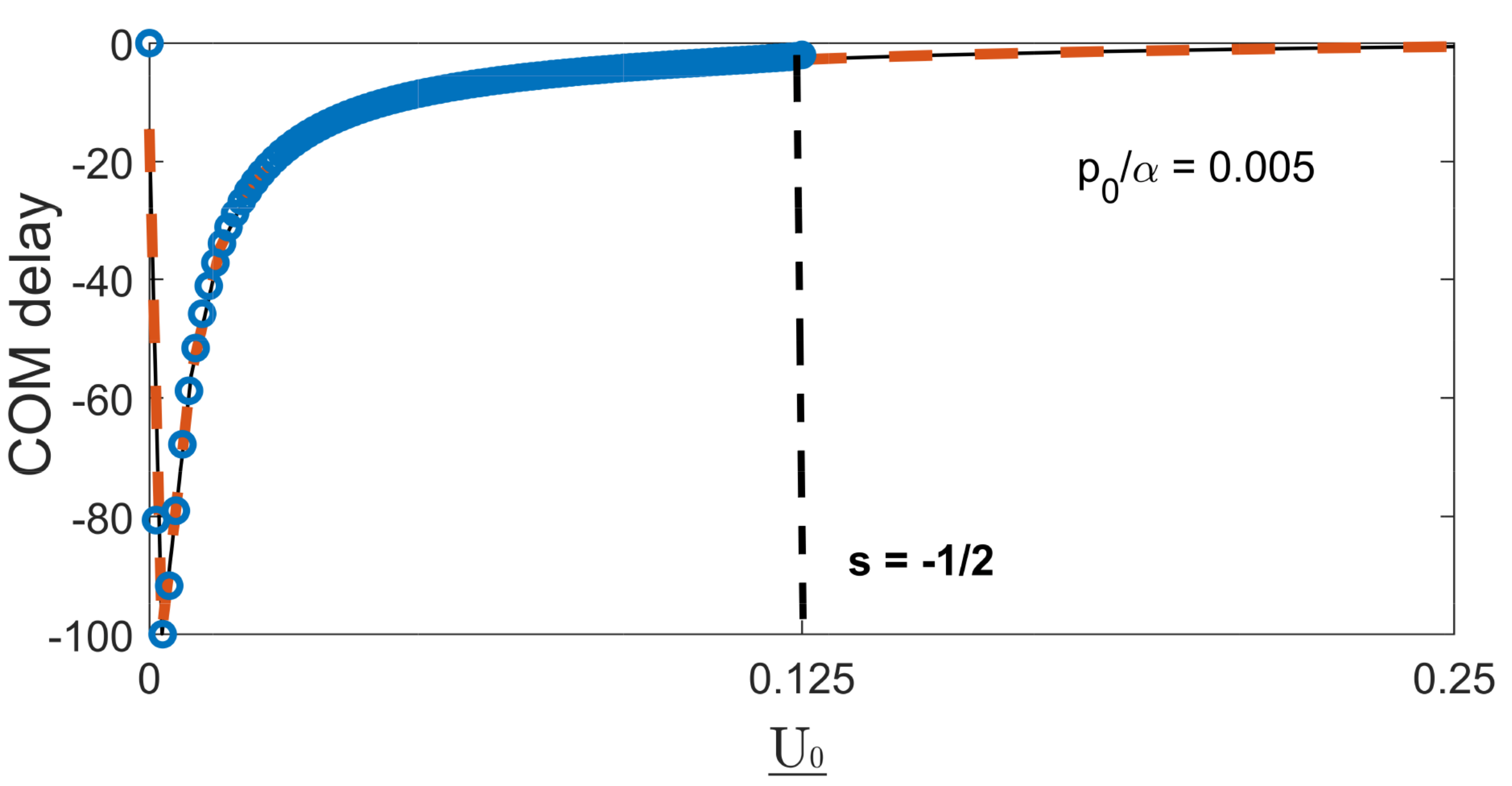}
\caption {Centre-of-mass delay of a wave packet with $\pp_0=0.005$, $\Delta \pp=0.001$, $\xx_0=-3*10^3$,  at
$\tt=2*10^6$ vs. the barrier's height $\uu > 0$,  calculated using equation (\ref{y1a})-(\ref{y1}) (solid), equation (\ref{f3}) (dashed), and equation (\ref{h3}) (dot-dashed)}
\label{7}
\end{figure}
Thus, a slow ($p_0/\alpha \ll 1$) and broad ($\alpha \Delta x \gg 1$) particle experiences the largest delay, $\delta x_{COM}- \delta v_0 t\approx 1/2p_0$,
is scattered by a barrier with $s \approx p_0/\al$.
This simple approximation begins to breaks down for $s\approx -1/2$ ($U_0\approx \al^2/8\mu$), where the poles of the fist and the second kind in Fig. 3 coalesce and, as the barrier increases, begin to move parallel to the real axis.
This change in the poles' behaviour has no visible effect  on either the transmission amplitude, or the delay in equation (\ref{y3}).
However, for $U_0 > \al^2/8\mu$, a larger number of poles must be taken into account in order to reproduce the transmitted
wave packet with sufficient accuracy, as shown in Fig. \ref{7}.
Note that the role of the poles lying far from the real axis is to cancel the contribution form the $\delta$-term in equation (\ref{c2})
and ensure the correct magnitude of the transmitted state (cf. Fig. 8).

\section{Conclusions and discussion}
In summary, there are two ways to look at transmission of a quantum particle across a short-range potential, be it a barrier or a well.
Firstly, the wave packet can be seen as probing the transmission amplitude, $T(p,V)$, in a range of momenta around its mean momentum, $p_0$, determined by the width of $A(p-p_0)$ in equation (\ref{1}). Integration in the momentum space gives the correct answer (\ref{2}) for the
transmitted wave packet, but provides little additional insight.
\newline
The second approach helps one to identify the disputed reshaping mechanism (see, e.g., \cite{{Kuritz},{Butt},{Winf}}).
equation (\ref{c1}) represents the transmitted state as a sum of freely propagating envelopes,
each shifted in space by a distance $x'$, and weighted by the corresponding probability amplitude
$\eta(p_0,x')$.
The problem, we argue,
is most naturally discussed in terms of the
particle's position at a given time \cite{FOOT}.
One recognises certain features familiar from classical mechanics.
For a barrier, all envelopes are delayed relative to free propagation, and it requires a potential well to have some of them
advanced. This does not, however, guarantee that the tunnelling particle will be found delayed if compared with a
freely
propagating one.
Since $\eta(p_0,x')$ may change sign, the centre of mass of the transmitted wave packet, constructed from the front tails
of the delayed envelopes, may end up advanced, as shown in Fig. 2{\bf c}.
\newline
One of the enduring misconceptions about the subject is the belief that the
time at which a
transmitted particle arrives at a fixed detector can be related
to the duration, spent by the particle in the barrier.
As discussed in Section X, the detection times are easily linked to the transmitted particle's instantaneous position.
{\crr Yet  they provide no further insight into which of the delays in Eq.(\ref{c2}) occurs in the barrier, and  for a very fundamental reason.}
With several spacial delays interfering, the Uncertainty Principle forbids identifying the one occurred in the same sense it leaves
indeterminate the slit chosen in a double slit experiment {\crr (see Supplementary Appendix E)}.
In general, there is no single spatial delay, associated with quantum transmission.
Worse still, since $\eta(p_0,x')$ changes sign, we found no simple way to characterise transmission by an effective range
of delays (see Sections IX and XII).
\newline
One exception is the classical limit.
For a classically allowed transmission, the amplitude distribution $\eta(p_0,x')$ selects a single
delay $\x'$, and a single  envelope  $G_0(x-\x', t)$, delayed or advanced relative to free propagation, as shown in Figs. 2{\bf a} and 2{\bf b}.
This is no longer true for semiclassical tunnelling, where $\eta(p_0,x')$ also has a saddle point,
but there the similarity ends. equation (\ref{d2}) sums the free envelopes along the real $x'$-axis, but the saddle at $\x' =\x'_1-i\x'_2$ lies
in the complex $x'$-plane and cannot be seen in Fig. 2{\bf f}. Interference between all real delays produces the resulting envelope,
$G_0(x-\x'_1-i\x'_2,t)$, which cannot be associated with a single real spatial shift.
\newline
Then what is the \e{phase time}, often associated with the duration of tunnelling \cite{Hauge}?
Firstly, it is what one obtains by dividing the distance separating the COM of a transmitted wave packet,
broad in the coordinate space, from the COM of its free counterpart, by the particle's mean velocity [cf. equation (\ref{f3a})].
Secondly, it can be expressed as the real part of  the first moment of  an alternating \e{distribution} $\eta(p_0,x')$ [cf. equation (\ref{f1})].
It can, therefore, be measured, but should not be interpreted as the \e{excess time spent in the barrier}
if one wishes to avoid a conflict with special relativity.
\newline
Furthermore, the transmitted state $\psi^T(x,t)$ can be written as a discrete sum over the singularities of the
transmission amplitude $T(p_0,V)$. This representation is convenient for describing low-energy scattering
by shallow wells or low barriers, as discussed in Sections XI and XII. It becomes impractical in the classical
or semiclassical limit, where individual terms of the pole sum become very large (see Fig. \ref{3}).
We note that a similar behaviour occurs in a different model (see Section 6 of \cite{DSE0}), using an
interference-based reshaping mechanism to advance the transmitted state.
\newline
In summary, we provided a detailed analysis of transmission across various Eckart potentials.
We conclude that, except in the classical limit, quantum scattering is essentially an interference phenomenon, not amenable to
simplistic descriptions in terms of  a single delay experienced in the potential, or even of a probability distribution of
such delays. This is the fundamental difficulty with the search for a \e{tunneling time} which began with McColl's
paper \cite{McColl} almost ninety years ago.

\appendix
\section*{Supplementary Information}
\section{Freely propagating envelopes and the error functions in Eq.(49)}

A freely propagating Gaussian envelope  $G_0(x,t)$ in equation
($8$) has the form
\begin{equation}\label{aa1}
G_0(x,t) =\left( \frac{2\Delta_x^2}{\pi\Delta_{x_t}^4} \right)^{1/4}
 \exp\left[ -(x- p_0t/\mu -x_0)^2/\Delta_{x_t}^2  \right] \tag{S1} 
\end{equation}
where $\Delta_{x_t}=\sqrt{\Delta x^2+2it/\mu}$. The integrals in equation 
($18$) can be expressed in terms of the error function \cite{Abr}, and we have
\begin{align}\label{gg1}
\Psi^T(x,t) = &\exp\left[ ip_0x-iE(p_0)t \right]\{ G_0(x,t)
 +\sum_{n_B} \text{Res}(k_{n_B}){\large\textfrak{G}^B}(x,k^I_n,p_0)
+\sum_{n_R} \text{Res}(k_{n_R}) {\large{\textfrak{G}^R}}(x,k^I_n,p_0)\}\tag{S2}
\end{align}
where
\begin{align}\label{aaer}
{\large\textfrak{G}}^{B,R}=\left( \frac{2\Delta_x^2}{\pi\Delta_{x_t}^4} \right)^{1/4}
\times \frac{\sqrt{\pi}\left[1 \pm  \text{erf}\left(b_{B,R}/2\sqrt{a}\right)\right]}{2\sqrt{a}}
\exp[b_{B,R}^2/4a + c]\tag{S3}
\end{align}
and
\begin{align}
&a = -1/\Delta_{x_t}^2, \q c = -(x-pt-x_0)^2/\Delta_{x_t}^2,\q\q\nonumber\\
&b_{B,R} = i(k_{n_{B,R}}-p_0) + 2(x-pt-x_0)/\Delta_{x_t}^2.\tag{S4}
\end{align}
For an Eckart well with $s=M$, we have equation 
($49$).
\section{Behaviour of $\et(p_0,x')$ for $x'\to 0$}
To see whether $\et(p_0,x')$ remains finite at $x'=0$, we note that using 
($6$) one can write the integral in equation 
($20$)
as
\begin{equation}\label{A1}
I \equiv \int_{-\infty}^\infty [T(k,V)-1]dk=\int_{0}^\infty \{\R[T(k,V)]-1\}dk.\tag{S5}
\end{equation}
Since $|T(k,V)|\le 0$, $\et(p_0,x=0)$ can only become infinite due to the behaviour of the integrand at large $k$.
As $k\to \infty$ the motion becomes semiclassical, and from
($10$) we have
\begin{equation}\label{A2}
I=\int_{0}^\infty  \{ \cos\left[ \Phi(k,V)\right]-1\}dk,\tag{S6}
\end{equation}
where for $\Phi(k,V)-1$ we find
\begin{equation}\label{A3}
 \Phi(k\to \infty ,V)-1\to (\mu/k)\int_{-\infty}^\infty V(x)dx + o(1/k).\tag{S7}
 \end{equation}
 With $J\equiv \int_{-\infty}^\infty V(x)dx $ finite, we have
 \begin{equation}\label{A4}
\cos[\Phi(k\to \infty,V)]-1 \to -\mu^2 J^2/2k^2,\tag{S8}
\end{equation}
integral (\ref{A2}) converges to a finite value, and we find $\et(p_0,x')< \infty$.
This is true for both barriers, $U_0>0$,  and wells, $U_0 <0$.
 \section{Double poles of the transmission amplitude $T(p,V)$ }
In a special case  $s = M  + 1/2$, $M=0,1,...$ on the imaginary axis there are $2M+2$ simple poles $k^I_n$, $n=0,1,...,2M+1$,
$M+1$ above, and $M+1$ below the real axis (cf. Fig. 3). The rest are double poles, since both Gamma functions in the numerator of equation 
($4$)
diverge at $p=k^I_n$, $n\ge 2M+2$. The corresponding residues can be obtained with the help of the Cauchy's differentiation formula 

\begin{align}\label{double_pole_res}
\text{Res}_2 & (k_n^I) =  \frac{2i\alpha}{n!(n+2M+2)!}
\times \left( \sum_{k = 1}^{k = n}\frac{1}{k} + \sum_{k=n+1}^{k = n+2M+2}\frac{1}{2k} - \gamma(1) \right)
\times \frac{1}{\Gamma(-n-M-3/2) \Gamma(-n-M-1/2)},\tag{S9}
\end{align}

where $\gamma(1)\approx 0.5772$ is the Euler-Mascheroni constant \cite{Abr}.
The final result, therefore, is

\begin{align}
\eta&(p_0,x^\prime) = \delta(x^\prime) + i \text{exp}(-ip_0x^\prime)
\times\begin{cases}
\sum_{n = 0}^M Res(k_n^I)\text{exp}(ik_n^Ix^\prime), \q \q\q\q  x^\prime \geq 0\\
\left [\sum_{n = M+1}^{2M+1} Res(k_n^I)\text{exp}(ik_n^Ix^\prime)\right. +\\
\left.  \sum_{n = 2M+2}^{\infty} Res_2(k_n^I)\text{exp}(ik_n^Ix^\prime)\right ], \q x^\prime < 0.\
\end{cases}\tag{S10}
\end{align}

\section{The residues in the limit  $n\to \infty$}
For a large $n\to \infty$ we have
\begin{equation}
\Gamma(z) \approx z^{z-1/2}e^{-z}\sqrt{2\pi},\tag{S11}
\end{equation}
which is valid for $|\arg(z)| < \pi$.
 Using the Sterling formula $n! \approx \sqrt{2 \pi n}\left( \frac{n}{e}\right)^n$ and applying this to the residues of the poles of the first kind in equation ($30$), yields
 \begin{align}
 &\text{Res}(k_n^I) \approx i\frac{\alpha (-1)^n}{2\pi\sqrt{n}}\times
 \frac{(-n+s)^{n-s+1/2}(-n+s+1)^{n-s-1/2}}{n^n(n-2s-1)^{-n+2s-1}}.\tag{S12}
 \end{align}
where
\begin{align}
\lim_{n\to\infty} & \frac{(-n+s)^{n-s+1/2}(-n+s+1)^{n-s-1/2}}{(n-2s-1)^{-n+2s-1}n^n}
=i(-1)^n\sqrt{n}.\tag{S13}
\end{align}
Recalling that  $\text{Res}(k_n^I) = -\text{Res}^*(k_n^{II})$, we have the desired limit
\begin{equation}
\lim_{n\to\infty} \text{Res}(k_n^I) = - \alpha/2\pi, \q
\lim_{n\to\infty} \text{Res}(k_n^{II}) =  \alpha/2\pi.\tag{S14}
\end{equation}
 \section{The double-slit analogy}
 To see what the double-slit conundrum and the problem at hand have in common, consider the simplest double-slit arrangement.
 This includes a two-level system with a Hamiltonian $\h_S$, prepared at $t=0$ is a state $|\psi\ra$ (the source), and 
 observed again in a final state $|\phi\ra$ (point on the screen). A pair of orthogonal state $|b_1\ra$ and $|b_2\ra$ in which the system could 
 be at $t=T/2$ play the role of the \e{slits}, so that the system can reach the $|\phi\ra$ by passing via $|b_1\ra$ or $|b_2\ra$, the corresponding amplitudes being [$\u_S(t)=\exp(-i\h_S)$]
\begin{align} \label{Dslit1}
A(\phi\gets b_n\gets  \psi)=\la \phi|\u_S(T/2)|b_n\ra \la b_n|\u_S(T/2)|\psi \ra\equiv \eta_n,\q n=1,2.\tag{S19}
\end{align}
The amplitude to arrive in $|\phi\ra$ results from the interference between the two alternatives, 
and the corresponding probability is $|\eta_1+\eta_2|^2$. 
The Uncertainty Principle (UP) \cite{FeynL} states that this \e{which way?} question {\it cannot be answered without destroying the interference}.
In order to determine the route taken by the system we can measure, at $t=T/2$, the \e{slit number operaror}
\begin{align}\label{Dslit2}
\hat{ \mathcal N}_S= 1\times |b_1\ra \la b_1|+2\times |b_2\ra \la b_2|, \tag{S20}
\end{align}
thus obtaining the result $n=1,2$, if the $n$-th route is taken. To do so we couple the system to a von Neumann pointer with position 
$x$, so that the full Hamiltonian becomes $\h=\h_S -i\partial_{x}\hat{ \mathcal N}_S$, and the initial state of the joint system is
$|\psi\ra \otimes |G\ra$, where the pointer's initial state can be chosen to be a real-valued Gaussian of a width $\Delta x$, centred at the origin, 
$\la x|G\ra= \la -x|G\ra$. Now the (unnormalised) probability to find the pointer at $x$, given that the system has arrived at $|\phi\ra$ is 
\begin{align}\label{Dslit3}
\rho(x)=|G(x-1)\eta_1+G(x-2)\eta_2|^2,\tag{S21} 
\end{align}
and everything depends on the accuracy of the measurement,  $\Delta x$. If $\Delta x <<1$, each trial produces an outcome
$1$ or $2$, we know where the system was at $t=T/2$, but the probability to arrive in $|\phi\ra$ has changed to $|\eta_1|^2+|\eta_2|^2$.
We destroyed the studied transition. 
\newline
To keep the transition more or less intact, we can try choosing a large $\Delta x$, $\Delta x \to \infty$.  Now
the probability of post-selection in $|\phi\ra$, and
 the pointer's readings may lie 
anywhere, $-\infty < x < \infty$. This agrees with the UP, which says that the route taken by the system cannot be determined in the presence 
of interference. We can, however, evaluate the {\it mean} pointer position which is easily found to be [cf. equation ($10$)] 
\begin{align}\label{Dslit4}
\la x\ra \equiv \int x \rho(x)dx/\int \rho(x)dx \approx \R\left[ \frac{1\times \eta_1+2\times \eta_2}{ \eta_1+ \eta_2}\right ]\q,\tag{S22} 
\end{align}
and treat $\la x\ra$ as the mean slit number, $\overline n$, in the presence of interference. The problem is that with complex values 
$\eta_n$, with no restrictions on the sign of either $\R[\eta_n]$ and $\Ip[\eta_n]$ there are also  no restrictions on the value of  $\la x\ra$. 
For example \cite{DSPLA} it is easy to find $|\psi\ra$, $|\phi\ra$, $|b_1\ra$ and $|b_2\ra$ for our mean slit number to be $100$. 
\newline
Can we measure $\overline n = \la x\ra =100$? Definitely yes. 
\newline
Do we really want to \e{explain} a situation, where we drilled only two holes in the screen, by saying that there are 
up to $100$ holes we did not know about? Most likely  not.
\newline
The same applies to the phase time $\tau_{phase}$ in equation 
($47$), where the particle's own position $x$ plays the role of the pointer's coordinate in (\ref{Dslit3}) \cite{DSE0}. 
As the width of the wave packet becomes very large, $\Delta x \to \infty$, for the position of the COM we have
\begin{align}\label{Dslit5}
x_{COM}=\la x\ra \equiv \int x |\psi(x,T)|^2dx/\int |\psi(x,T)|^2dx
\approx \R\left[ 
\frac{\int x'\eta(x')dx'}{\int \eta(x')dx'}
\right]\tag{S23}
\end{align}
When this is used to deduce the value of $\tau_{phase}$, the value turns out to be very small. 
\newline
Can we measure this short duration? Definitely yes \cite{Nat}.
\newline 
Do we really want to claim that a tunnelling particle defies relativity by moving too fast in the barrier, when all the barrier can do is delay it?
Most likely not (with few exceptions \cite{Nimtz}).

So what is the meaning of the (measured) values (\ref{Dslit4}) and (\ref{Dslit5})? They express the correct relations between 
Feynman's transition amplitude \cite{FeynL}, which quantum mechanics uses to describe the phenomena in question, 
and very little else \cite{DSPLA}.
 \section{The centre-of-mass delay for $s\approx M$, $M=1,2,3...$} 
 For $s \approx M$, where the $M$-th bound state enters the well as $s$ increases, we have
\begin{align}
\tilde{\eta}&(p,x^\prime) \approx \delta(x^\prime) + \sum_{n = 0}^{M-1} i\ \frac{\alpha (-1)^n(2M-n)!}{n!(M-n-1)!(M-n)!}
\times\exp \{ -\left[ \alpha(M-n) + ip \right]x^\prime\}\theta(x')\nonumber\\
&+i \frac{\alpha(-1)^M(2M-1)!}{M!(M-1)!}(s-M) \exp\{ -\left[\alpha(s-M)+ip \right]x^\prime \}
\times [\theta(x')\theta(s-M)+
\theta(-x')\theta(M-s)],\tag{S15}
\end{align}
and
\begin{align}
T(p,V) \approx \exp[i\Theta(p)]
- \frac{(-1)^M(2M-1)!}{M!(M-1)!}
\times\frac{(s-M)}{(s-M)+ip/\alpha},\tag{S16}
\end{align}
where
\begin{align}
\Theta(p)=-i\ln \left \{1-
\sum_{n = 0}^{M-1} \frac{(-1)^n(2M-n)!}{n!(M-n-1)!(M-n)!}\right.
\left.\times \frac{1}{(M-n)+ip/\alpha}\right \}.\tag{S17}
\end{align}
The centre-of-mass delay, corrected for momentum filtering, is, therefore, given by
\begin{equation}
\delta x_{COM}^T(t) -\delta v_0 t
\approx
\frac{\alpha(s-M)}{\alpha^2(s-M)^2+p_0^2}-\partial_p\Theta(p_0).\tag{S18}
\end{equation}


\section*{Acknowledgements}
Financial support through the grants
PGC2018-101355-B-100 funded by MCIN/AEI/ 10.13039/501100011033 and by “ERDF A way of making Europe”, PID2019-107609GB-I00 by MCIN, and the Basque Government Grant No IT986-16,
is acknowledged by MP and DS.


\begin{thebibliography}{10}
\bibitem{FOOT} One can, in principle, consider a Fourier transform with respect to the energy $E$,
and represent the transmitted state as a weighted sum of the free envelopes, shifted in time rather than in space.
The analyses is the same if the dispersion law is linear, $E=cp$, but becomes more complicated for massive particles,
$E=p^2/2\mu$ \cite{DSE1}, since $T(E,V)$ is no longer single-valued in the complex energy plane.
For this reason we prefer an analysis in terms of spacial delays.
\end{thebibliography}
\end{document}